\documentclass[10pt, journal, letterpaper, twocolumn]{IEEEtran}

\ifCLASSOPTIONcompsoc
\usepackage[nocompress]{cite}
\else
\usepackage{cite}
\fi

\ifCLASSINFOpdf

\else

\fi
\usepackage{booktabs}

\usepackage{amssymb}
\usepackage{pifont}
\newcommand{\cmark}{\ding{51}}%
\newcommand{\xmark}{\ding{55}}%

\usepackage{amsmath}
\usepackage{color}
\usepackage[table,xcdraw]{xcolor}
\usepackage{graphicx,booktabs}
\usepackage{rotating,multirow,multicol}
\usepackage{array}
\usepackage{blindtext}
\usepackage{amssymb}%
\usepackage{pifont}%
\usepackage{caption}
\usepackage{subcaption}
\usepackage{soul}
\usepackage[normalem]{ulem}

\definecolor{maroon}{cmyk}{0,0.87,0.68,0.32}
\definecolor{LightCyan}{RGB}{155, 227, 247}
\usepackage[british,english,american]{babel}
\usepackage{color, colortbl}
\definecolor{Gray}{gray}{0.9}

\usepackage{enumitem}

\usepackage{adjustbox}

\usepackage{balance}
\usepackage{siunitx}
\usepackage{hyperref}

\begin{document}
\title{\huge Online Advertising Security: \\Issues, Taxonomy, and Future Directions}

\author{Zahra~Pooranian,~\IEEEmembership{Senior~Member,~IEEE,}~Mauro~Conti,~\IEEEmembership{Senior~Member,~IEEE},\\Hamed Haddadi,~\IEEEmembership{Member,~IEEE}~and~Rahim~Tafazolli,~\IEEEmembership{Senior~Member,~IEEE}

	\IEEEcompsocitemizethanks{\IEEEcompsocthanksitem Zahra Pooranian and Rahim Tafazolli are with 5G \& 6G Innovation Centre (5GIC \& 6GIC), Institute for Communication Systems (ICS), University of Surrey, Guildford, UK \protect (e-mail: \{z.pooranian, r.tafazolli\}@surrey.ac.uk) 	
		\IEEEcompsocthanksitem Mauro Conti is with the Department of Mathematics, University of Padua, Padua, Italy \protect (e-mail: conti@math.unipd.it) 
		\IEEEcompsocthanksitem Hamed Haddadi is with the Faculty of Engineering, Imperial College London, London, UK \protect (e-mail: h.haddadi@imperial.ac.uk)}
	
	\thanks{Manuscript received 19 May 2020; revised 6 Dec 2020, 17 May and 16 July 2021; accepted 2 October 2021}}

\markboth{IEEE Communications Surveys \& Tutorials,~Vol.~XX, No.~NN,~XX~2021}{Zahra Pooranian \MakeLowercase{\textit{et al.}}: Online Advertising Security: Issues, taxonomy, and future directions}

	\maketitle	
	\begin{abstract}
		
Online advertising has become the backbone of the Internet economy by revolutionizing business marketing. It provides a simple and efficient way for advertisers to display their advertisements to specific individual users, and over the last couple of years has contributed to an explosion in the income stream for several web-based businesses. For example, Google's income from advertising grew 51.6\% between 2016 and 2018, to \$136.8 billion. This exponential growth in advertising revenue has motivated fraudsters to exploit the weaknesses of the online advertising model to make money, and researchers to discover new security vulnerabilities in the model, to propose countermeasures and to forecast future trends in research. 

Motivated by these considerations, this paper presents a comprehensive review of the security threats to online advertising systems. We begin by introducing the motivation for online advertising system, explain how it differs from traditional advertising networks, introduce terminology, and define the current online advertising architecture. We then devise a comprehensive taxonomy of attacks on online advertising to raise awareness among researchers about the vulnerabilities of online advertising ecosystem. We discuss the limitations and effectiveness of the countermeasures that have been developed to secure entities in the advertising ecosystem against these attacks. To complete our work, we identify some open issues and outline some possible directions for future research towards improving security methods for online advertising systems.

\end{abstract}


\maketitle	


\normalsize

\section{Introduction}
\label{sec:Intro}

\IEEEPARstart{O}{ver} the past few years, the widespread adoption of the Internet has led to the emergence of a new form of online business -- i.e., \emph{online advertising} -- to make money through this means. A significant financial pillar of the Internet ecosystem is provided by online advertising (from websites and mobile apps)~\cite{vratonjic2008securing, crussell2014madfraud, song2013multi, gill2013best}.

Many companies such as Google and Microsoft have increased their investment in online advertising to improve their revenue and sales. According to the report in~\cite{Forbes}, Google's income from advertising grew 51.6\% between 2016 and 2018, to \$136.8 billion. It was expected that this revenue reached nearly \$203.4 billion by 2020 and will continue to increase over time. Also, mobile advertising has become one of the fastest-growing industries with the advent of smartphones~\cite{yao2020botspot}. Millions of mobile applications are registered in various application platforms such as Google Play Store, Apps Store, etc., which contain at least one advertising library that allows mobile advertising~\cite{ullah2020privacy}. According to~\cite{eMarketer}, in 2019, total mobile advertising spending worldwide has reached \$189 billion and will surpass \$240 billion by 2022.

\begin{figure}
	\centering
	\includegraphics[width=0.50\textwidth]{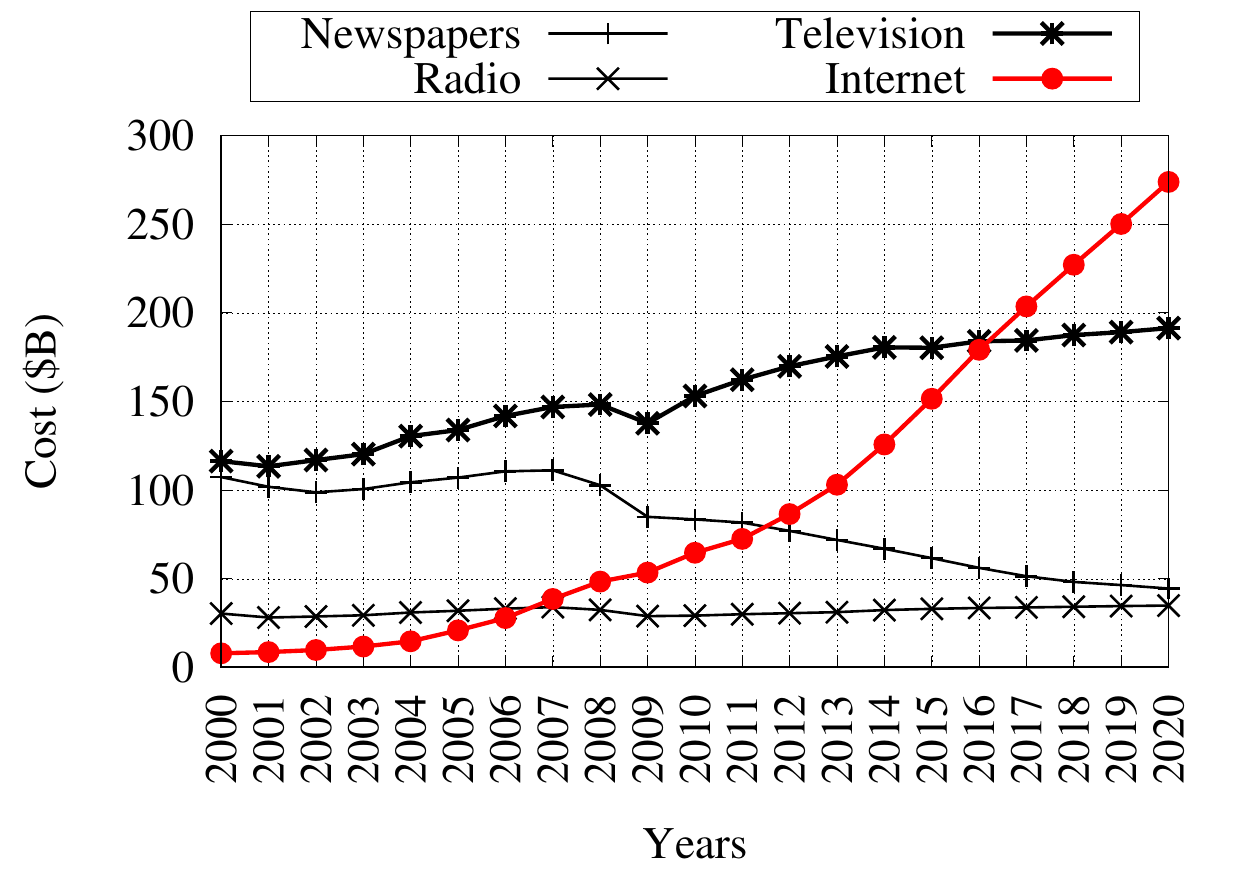}
	\caption{\small Global ad spending by medium.}\label{fig:adspending}
\end{figure}

Online advertising uses the same mechanisms that are applied to manage other ``traditional'' advertising channels, such as newspapers, radio or TV, but is much more creative in providing targeted and personalized advertisements~\cite{yurovskiy2015pros, haddadi2011targeted}. Thanks to the rise of the Internet and online advertising, sales of TV and radio advertisements have stagnated, and those of newspaper advertisements have dropped. Fig.~\ref{fig:adspending} shows a comparison of global ad spending by medium~\cite{Recode}.

Online advertising provides profit for all the components of the system, such as publishers, advertisers, and advertising network (ad network). Given the high profits involved, the online advertising system is an obvious target for fraud. Hence, several attacks on the current online advertising market have been identified that have targeted various entities in the market, such as hacking campaign account~\cite{mladenow2015online}, click fraud~\cite{linden2012method}, inflight modification of advertising (ad) traffic~\cite{vratonjic2011online}, and malvertising~\cite{poornachandran2017demalvertising}.




The inherent lack of transparency and complexity of the online advertising ecosystem give rise to higher risks, and an adversary can easily exploit these aspects to engage in fraudulent activities and launch an attack on the system. Ad fraud can occur in various forms and may involve fooling different components of the online advertising ecosystem to make money. For instance, dishonest publishers may deceive advertisers into paying an extra fee, or hackers could hijack an advertising slot to gain revenue for themselves.

In view of the factors described above, the success and popularity of the online advertising ecosystem depend primarily on the level of security that can provide against such malicious threats.
The considerations mentioned above motivate the current work in terms of studying security issues in the online advertising market and essential related techniques.

\textbf{Survey Organization:} The remainder of this article is structured as follows. We begin by discussing work that gives a high-level overview of related online advertising survey, which help define the contributions of this paper (Section~\ref{sec:RelatedSurvey}). In Section~\ref{sec:background}, we explain the differences between the online advertising system and traditional advertising networks, introduce the terminology used, and describe its architecture. Section~\ref{sec:attack_adv} presents our proposed taxonomy of vulnerability on the online advertising system. We also discuss the goal of these attacks, the revenue model, and the primary components targets. In Section~\ref{sec:count_mea}, we categorize and discuss various security solutions identified in the literature and present a preliminary overview of the advantages and disadvantages of the use of these solutions in online advertising systems. In Section~\ref{sec:future}, we highlight several open challenges for future research in online advertising systems. We conclude our work in Section~\ref{sec:coclu}.

\section{Related High Level Articles and The Scope of This Survey}
\label{sec:RelatedSurvey}

We begin this section by discussing the related online advertising survey in Section~\ref{sec:history}, which help define the scope and contributions of this paper in Section~\ref{sec:Cont}.

\subsection{Surveys on Online Advertising Systems}
\label{sec:history}

Numerous existing works have discussed general aspects of online advertising systems. Most of the early works focused on issues relating to the economic aspects of advertising~\cite{edelman2007internet},~\cite{evans2008economics},~\cite{tucker2012economics},~\cite{chen2014economic},~\cite{evans2009online}, the literature review on online advertising~\cite{choi2020online}, challenges in online advertising~\cite{bostanshirin2014online}, theoretical or analytical assessments of sponsored searches~\cite{aggarwal2008sponsored},~\cite{goldfarb2008search},~\cite{varian2007position}, and especially analyses of privacy threats and protection mechanisms~\cite{estrada2017online},~\cite{chen2016depth}.

The authors of~\cite{dong2018frauddroid} investigated a wide range of mobile ad frauds and developed a comprehensive taxonomy for the research community. Budak et al.~\cite{budak2016understanding} showed that one of the leading sources of threats to the online advertising system is the widespread use of ad-blocking software and third-party platform tracking. The study in~\cite{ullah2020privacy} presented a comprehensive survey of the existing literature on the privacy risks of targeted advertising, together with solutions to these challenges. The papers in~\cite{kayalvizhi2018survey}~and~\cite{gohilsurvey} surveyed click fraud attacks and analyzed solutions for mitigating them, with a focus on the threats affecting the PCC revenue model. In~\cite{daswani2008online}, the authors provided a summary of the current state of click fraud, including typical forms of attack and their countermeasures. Although they discussed the impact of click fraud on different types of revenue models, less attention was paid to other kinds of ad fraud in online advertising systems. The survey in~\cite{laleh2009taxonomy} gave a detailed summary of all the forms of web and network fraud that can be detected by data mining techniques, and reviewed some approaches for identifying fraud in real time. The study in~\cite{andvaleriodigital} focused on the visualization of the online advertising ecosystem and future trends in programmatic advertising; although the authors aimed to address a gap in the literature by developing a standard for the visualization and explanation of the digital advertising ecosystem, the survey did not include any discussion of the security aspects of online advertising systems. The survey in~\cite{cai2020threats} addressed the security problems faced by online advertising, including the corresponding countermeasures in the literature. However, our survey paper covers online advertising systems from various angles as follows. First, we examine threats in the system from three dimensions, depending on whether the page content is targeted by fraud, ad traffic, or user actions. Therefore, we classify advertising fraud into three main categories: placement fraud, traffic fraud, and action fraud. In particular, we explain how an adversary can exploit the risks in online advertising systems and conduct ad fraud for each type of attack. Second, we provide a detailed discussion of the goals of the attacks, the revenue model that is the attacker's target and the primary component targets (see Table~\ref{table:tbl_attack}). Third, we have provided the pros and cons of the countermeasures techniques and how they could combat attacks. Finally, in the future section, we have discussed four major aspects of research to support online advertising systems' security, reliability, and efficiency.

\subsection{Our Scope}
\label{sec:Cont}
This article presents a survey that primarily targets the security issues and challenges of online advertising systems and reviews the related fundamental concepts. From a security perspective, it presents a comprehensive taxonomy of well-known ad fraud. It also categorizes several security mechanisms that have been proposed in recent years to cope with and mitigate the existing security challenges in the online advertising industry. In particular, our classification focuses on the goals of attacks, the revenue model, and the primary component targets.


The research papers and books we mentioned previously did not address security issues with an emphasis on online ad fraud in this area. There is therefore a need for a concise survey to provide a reader who is planning to undertake research in this field with a classification of online ad fraud, along with an exhaustive review of the corresponding countermeasures. In brief, the essential contributions of the survey are as follows: 

\begin{itemize}

	\item First, some essential background knowledge is presented, including the differences between traditional and current online advertising systems, the terminology used, and the existing architecture of online advertising (e.g., web and mobile). The goal is to enable new readers to gain the required familiarity with online advertising systems and its underlying technologies, such as revenue models and the payment of commissions.
	
	\item We present a detailed taxonomy of the current security threats to online advertising. We investigate several possibilities, including both theoretical and practical vulnerabilities, that fraudsters can use to launch an attack on the online advertising industry. In addition, we present a detailed discussion of the goals of these attacks, their impact on the particular revenue models, and the primary component targets. 
	
	\item We review several cutting-edge solutions that address security threats to online advertising systems, and explain the advantages and disadvantages of each solution.
	
	\item Finally, we identify a number of open challenges and future research directions in the field of online advertising, with particular attention to the security aspects.
	
\end{itemize}

To the best of our knowledge, there are no existing surveys that have reviewed and summarized the existing security vulnerabilities and outlined future research directions in the realm of online advertising systems. Motivated by this consideration, the main goals of this study are threefold: $(i)$ to help the reader to understand the scope and consequences of the security threats and challenges in the domain of online advertising systems; $(ii)$ to estimate the potential damage associated with these threats; and $(iii)$ to highlight paths that are likely to lead to the detection and containment of these threats. From a practical perspective, our research aims to raise awareness in the online advertising research community of the urgent need to prevent various attacks from disrupting the healthy online advertising market.

\section{Overview of Online Advertising Systems}
\label{sec:background}



We begin this section with an explanation of the most widely used terminology associated with the online advertising ecosystem, including terms used in the remainder of the present article in Section~\ref{sec:term}. Section~\ref{sec:online_adv} provides a brief introduction to the online advertising ecosystem, especially in terms of its main components, the interactions between them, and the technologies they support. To gain insight into how online advertising networks operate, we discuss the current ads delivery workflow on the web in Section~\ref{sec:ad_online}. We highlight the similarities and differences between mobile advertising and web-based advertising ecosystems in Section~\ref{sec:mobile_adv}. Then, we discuss the ad delivery workflow on mobile platforms in Section~\ref{sec:ad_mobile}. Next, different methods of targeted advertising and the most common types of ad campaign (revenue model) are described in Sections~\ref{sec:targetAd} and~\ref{sec:revenue}, respectively. Finally, we explore how advertisers pay commission fees to commissioners and publishers in Section~\ref{sec:pay_commi}, and describe the main security goals in the online advertising system in Section~\ref{sec:SecurityR}.


For ease of reading, in Table~\ref{table:tbl_Symb}, we list the all abbreviations used in this paper.

	\begin{table}[]
		\caption{\small List of abbreviations and corresponding
			descriptions.}
		\label{table:tbl_Symb}
		\centering
	\begin{tabular}{|c|p{6.3cm}|} 
			
		\hline
		\textbf{Abbreviation} & \textbf{Description}\\
		\hline
		\rowcolor[HTML]{EFEFEF}
		Ad Network  & Advertising Network \\ \hline 
		Ad & Advertising  \\ \hline
		\rowcolor[HTML]{EFEFEF} 
		DSPs & Demand-side Platforms \\ \hline
		SSPs  & Supply-side Platforms \\ \hline
		\rowcolor[HTML]{EFEFEF} 
		HTTP & Hypertext Transfer Protocol \\ \hline
		RTB & Real-time Bidding \\ \hline
		\rowcolor[HTML]{EFEFEF} 
		CTR & Click-Through Rate \\ \hline
		ROI & Return on investment \\ \hline
		\rowcolor[HTML]{EFEFEF} 
		CM & Cookie Matching \\ \hline
		SDK & Software Development Kit \\ \hline
		\rowcolor[HTML]{EFEFEF} 
		UI & User Interfaces  \\ \hline
		GDPR & General Data Protection Regulation \\ \hline
		\rowcolor[HTML]{EFEFEF} 
		CPM & Cost per Impression Mile \\ \hline
		CPC & Cost per Click \\ \hline
		\rowcolor[HTML]{EFEFEF} 
		CPA & Cost per Action \\ \hline
		CIA & Confidentiality, Integrity and Availability \\ \hline
		\rowcolor[HTML]{EFEFEF} 
		ISP & Internet Service Provider \\ \hline
		MITM & Man-In-The-Middle \\ \hline
		\rowcolor[HTML]{EFEFEF} 
		DNS & Domain Name System \\ \hline
		CFC & Click Fraud Crowdsourcing \\ \hline
		\rowcolor[HTML]{EFEFEF} 
		MTA & Mail Transfer Agent \\ \hline
		TTP & Trusted Third Party \\ \hline
		\rowcolor[HTML]{EFEFEF} 
		CGI & Common Gateway Interface \\ \hline
		GBF & Group Blooms Filter \\ \hline
		\rowcolor[HTML]{EFEFEF} 
		TBF & Timing Blooms Filter \\ \hline
		NMF & Non-negative Matrix Factorization\\ \hline
		\rowcolor[HTML]{EFEFEF} 
		SeqGAN & Sequence Generative Adversarial Generative \\ \hline
		MLE & Maximum Likelihood Estimation\\ \hline
		\rowcolor[HTML]{EFEFEF} 
		CSBPNN & Cost-sensitive Back Propagation Neural Network\\ \hline
		ABC & Artificial Bee Colony\\ \hline
		\rowcolor[HTML]{EFEFEF} 	
		SLEUTH & Single-publisher attack dEtection Using correlaTion Hunting \\ \hline 
	    ML & Machine Learning\\ \hline 
	    \rowcolor[HTML]{EFEFEF} 
	    CFXGB & Cascaded Forest and XGBoost\\ \hline 
		CAPTCHA & Completely Automated Public Turing test to tell Computers and Humans Apart \\ \hline
		\rowcolor[HTML]{EFEFEF} 
		HMSM & Hidden Markov Scoring Model\\ \hline
		HMM & Hidden Markov Model \\ \hline 
		\rowcolor[HTML]{EFEFEF} 
			SVM & Support Vector Machine \\ \hline 
			TLS & Transport Layer Security\\ \hline 
			\rowcolor[HTML]{EFEFEF} 
		IoT &  Internet of Things\\ \hline 
			AI &  Artificial Intelligence\\ \hline
			\rowcolor[HTML]{EFEFEF} 
			AR & Augmented Reality\\ \hline
			5G & 5\textsuperscript{th} Generation of Mobile Internet\\ \hline
			\rowcolor[HTML]{EFEFEF} 
			DLT & Distributed Ledger Technology\\\hline
			API & Application Programming Interface\\
		\hline
	\end{tabular}
\end{table}

\subsection{Terminology}
\label{sec:term}

In this subsection, we define some essential terminology related to online advertising ecosystems, as used throughout this paper.

\begin{itemize}[leftmargin=*]
	\item{An \emph{advertiser} is a party who is willing to show a product, service, or event to the user via advertisements, in order to promote sales or attendance. Advertisers typically pay (or buy traffic from) an ad network to display their advertisements in the advertising space on publishers' websites or phone applications. The publisher also receives a percentage of this fee.}

	 \item{A \emph{publisher} (such as The New York Times or CNN) is an entity that receives money (via selling traffic) from advertisers by displaying their advertisements to users through its web pages (or mobile app).} 

	\item{A \emph{user} is an individual who visits a publisher's web pages.}

	\item{An \emph{advertising network} (such as Google, Yahoo, Google AdSense, Media.net, or PulsePoint) also known as a \emph{commissioner}, is part of an ad exchange. It acts as a broker between the advertiser and the publisher to manage the interaction between them~\cite{blizard2012click}, and is responsible for finding suitable spaces to present advertisements on publishers' websites for advertisers. They may also buy or sell ad traffic (as ad requests), either internally or together with other ad networks.}

	\item{An \emph{ad exchange} (such as DoubleClick~\cite{doubleclick}, AdECN~\cite{AdECN}, or OpenX~\cite{OpenX}) is a graph of the advertising networks that allows the advertiser and publisher to serve advertisements more effectively within an advertising space.}
	
	\item{\emph{Demand-side platforms (DSPs)} are components that work for advertisers, that is, for the actors who generate the demand for advertising services~\cite{gohilsurvey}. DSPs work on behalf of advertisers in front of ad exchanges, helping advertisers choose the right audience and media to display their ads. By gathering demand, DSPs can increase selection and effectiveness for advertisers.}
	
	\item{\emph{Supply-side platforms (SSPs)} act on behalf of publishers to provide advertising space to advertisers. SSP offers publishers an optimized strategy for managing their ad inventory.}
	
	\item{\emph{Ad servers} are a type of web server (or platform) that is used to host the content of an online advertisement and distribute this content on digital platforms such as Facebook, Quora, Twitter, etc.}
	
	\item{An \emph{advertising request} (ad request) is a query, in the form of Hypertext Transfer Protocol (HTTP) traffic, that is triggered by a web user's impressions or clicks, and calls an ad server to display an ad to the user.}

	
	\item{\emph{Creative} content is associated with the actual advertising message (e.g., an anchor tag, an Adobe Flash animation, text, or images) in the ad slot displayed to the user. The process of linking an ad message to an advertiser's website is called \emph{click-through}~\cite{stone2011understanding}.}

	\item{An ad server enumerates a \emph{click} event when a user clicks on an ad.}
	
	\item{An ad server counts an \emph{impression} event whenever the content or ad page is loaded for the user. Clicks and impressions generate two different events, which are handled separately in the online advertising system.}
	
	\item{An \emph{auction} is a competitive process that runs within the ad exchange. It is designed to allow each advertiser to bid for advertisement space, where the highest bidder is permitted to place an advertisement in the slot. An auction aims to generate more profit for publishers. In general, the time taken to complete the entire process is on the order of 100 ms.}

	\item{After an auction, ad networks may perform \emph{arbitrage} to increase their revenue. To initiate arbitrage, the ad network must run a new and independent auction by buying and reselling traffic from the publisher.}

	\item{An \emph{ad campaign} is a method that emerged to help advertisers to decide how much to pay when their advertisements are displayed. We discuss the most common forms of ad campaigns in Section~\ref{sec:revenue}.}

	\item{A \emph{banner} is a space on a page that displays a message from the advertiser.}
	
		\item{\emph{Real-time bidding} (RTB) is one of the critical technologies to make a profit for online advertising and allows advertisers to compete in real-time auctions to display their ads~\cite{estrada2019regulation}. Therefore, when a user visits a website, his impression is sold to the advertiser (or DSP), which offers the highest price in a few milliseconds. Moreover, the bid request messages received by DSPs contain user information (tracking data) to help them adjust ads based on user preferences and decide on a bidding strategy. In this way, the goal of RTB is twofold: to provide a personalized experience to users through targeted advertising and to maximize the profits of the entire advertising ecosystem.}

\end{itemize}

\subsection{An Overview of Web Advertising Ecosystem}
\label{sec:online_adv}

Not surprisingly, advertising techniques have evolved over time with the growth of the Internet, and online advertising has become one of the biggest and most profitable Internet businesses. The main idea behind online advertising is to provide an advertiser with a cost-effective, easy, fast, and flexible way to promote and sell their products through the Internet to suitable customers. In this way, it can maximize revenue, click-through rate (CTR\footnote[1]{ The CTR is the number of clicks an advertiser (i.e., publisher or ad) gains as a proportion of the impressions~\cite{jiang546multi}.}) or return on investment (ROI\footnote[2]{ ROI is an indicator used to measure the efficiency of an investment.}) of the advertising campaign~\cite{zhao2019deep}. There are several significant differences between current online advertising and traditional advertising (e.g., via television, radio, and newspapers). For example, traditional advertising uses massive broadcast advertisements without considering the user's interests; in contrast, online advertising can deliver advertisements to targeted users based on their interests and browsing behavior, regardless of geographical barriers.

%

\begin{figure}
	\centering
	\includegraphics[width=0.50\textwidth]{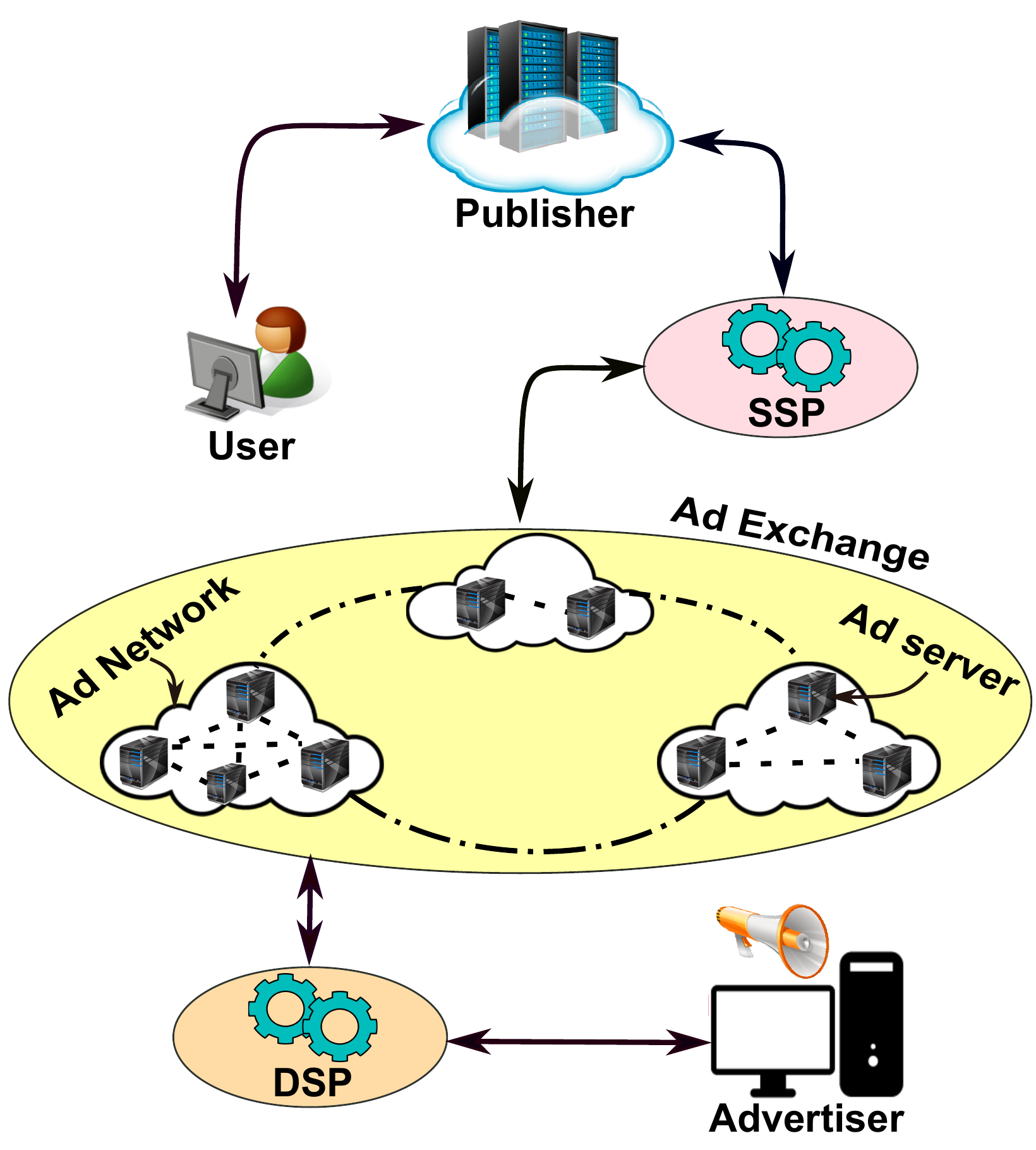}
	\caption{\small High-level overview of online advertising ecosystem.}\label{fig:ad_arch}
\end{figure}

Fig.~\ref{fig:ad_arch} shows the high-level overview of an online advertising ecosystem. Existing architecture can be more complex and dynamic than this design. However, the scheme relies on integrating three main components: an advertiser, a publisher, an ad exchange (e.g., multiple ad networks). Thanks to technologies such as RTB, the ultimate goal of these components is to show the right ad to the right user at the right time. The former two components show the demand and supply aspects of an online advertising service's economic model. The interaction between such players is usually done by an intermediate infrastructure, called ad exchange. To participate in the ad bidding, publishers and advertisers connect to the ad exchange network via SSPs and DSPs to conduct auctions and manage bids, respectively, then finally deliver the ads to various media platforms (such as a third party website, a search engine results page)~\cite{gharibshah2021user}. Users whose data and requests are the basis of decisions made for online advertising services are a passive part of this infrastructure because they do not make money from this billion-dollar business.


\subsection{General Operation of Web Advertising} \label{sec:ad_online}

By showing the main components of the online advertising ecosystem, we now give a brief description of how ads are delivered on the web.

\begin{figure}
	\centering
	\includegraphics[width=0.5\textwidth]{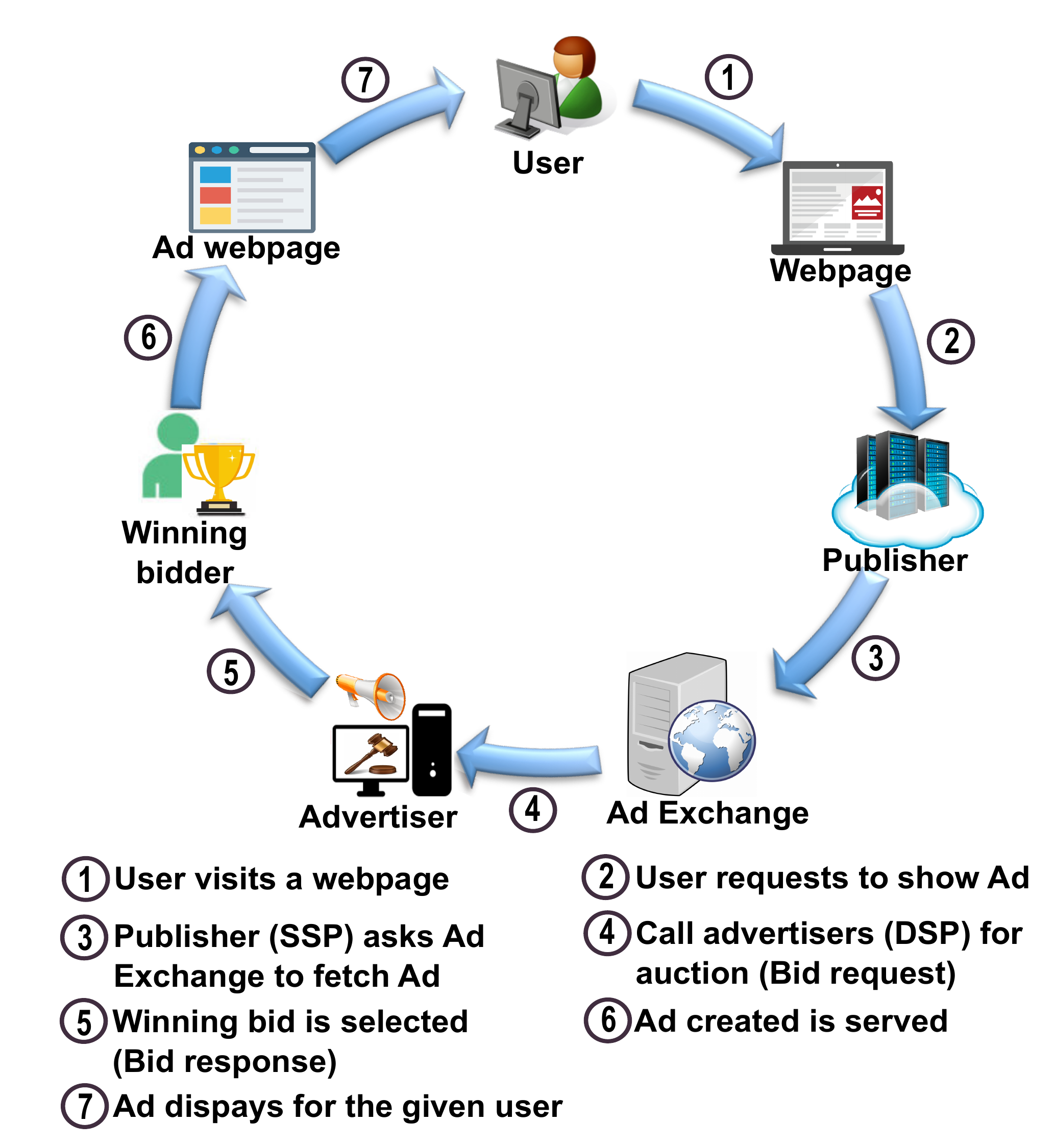}
	\caption{\small The process of serving ads in an online advertising system.}\label{fig:auction}
\end{figure}

The process of ad serving in an online advertising system is illustrated in Fig.~\ref{fig:auction}. The process is initiated when a user requests (e.g., an HTTP request) calls for an advertisement to be served by the publisher (step~1-2). Following this, the publisher (with the help of SSPs) asks the ad exchange to fill the ad on the visited page that best matches the user's profile and has the best price (step~3). The ad exchange starts an auction between multiple advertisers (with the help of DSPs) by sending the ``\emph{bid request}'' (with user data) to determine which can make the most profit for the publisher and, consequently, the whole network~\cite{niu2017era} (step~4). Both RTB and cookie matching (CM) mechanisms help the online advertising system ensure the most impact on users (which is to the benefit of advertisers), with the most benefit to publishers. RTB enables advertisers to bid for the chance to display an ad on a web page loaded by a user's browser.
	
Along with the bid request, ad exchanges send the following data about the user: URL of the page visited by the user; page subject category; IP address of the user or parts of it; and other information about their web browser. Ad exchanges widely use cookies and advertisers to collect and share such information, thus improving the advertising targeting process's accuracy~\cite{ghosh2015match}. Also, advertisers can use cookies to build a user's profile with information about their purchasing habits and browsing history for a future auction. Such information, through CM technology, helps advertisers and DSPs decide whether and how much to bid for a click or an impression. If an advertiser interested to show their ad, then send the price to the ad exchange. After such a process, the highest paying advertiser (winning bidder) wins the auction (step~5), and its ad is served and displayed to the user (steps~6-7).
 

\subsection{An Overview of Mobile Advertising Ecosystem}
\label{sec:mobile_adv}

Mobile advertising uses many traditional web advertising infrastructures, such as (mobile) user, (app) publisher, ad network, and advertiser, to deliver ads. An advertiser has the exact role of advertising in web advertising. An ad network is a trusted intermediary platform between the advertiser and app publisher to manage them. An app publisher (also called a developer) is an entity that publishes apps in the app market. The mobile user downloads applications from the application market and uses them on smartphones. Also, in this ecosystem, SSP and SSP are connected through an ad network.

In both web and mobile advertising systems, the ad library receives content embedded in a webpage or mobile app from ad providers and displays it on a webpage or mobile app interface. Then, the ad provider pays the publisher based on the number of clicks and impressions by the user. Despite the similarities, there are differences. For example, a website includes JavaScript code for displaying ads, while a mobile application has a custom software development kit (SDK\footnote[3]{ The SDK usually consists of a pre-compiled ad library with required dependencies~\cite{jin2019madlens}.}) embedded for loading JavaScript code and ads in a particular component like WebView\footnote[4]{ WebView displays web content directly in Android applications without directing the user to the web browser.} in Android~\cite{imamura2021web}.

	
\subsection{General Operation of Mobile Advertising}\label{sec:ad_mobile}

\begin{figure}
	\centering
	\includegraphics[width=0.5\textwidth]{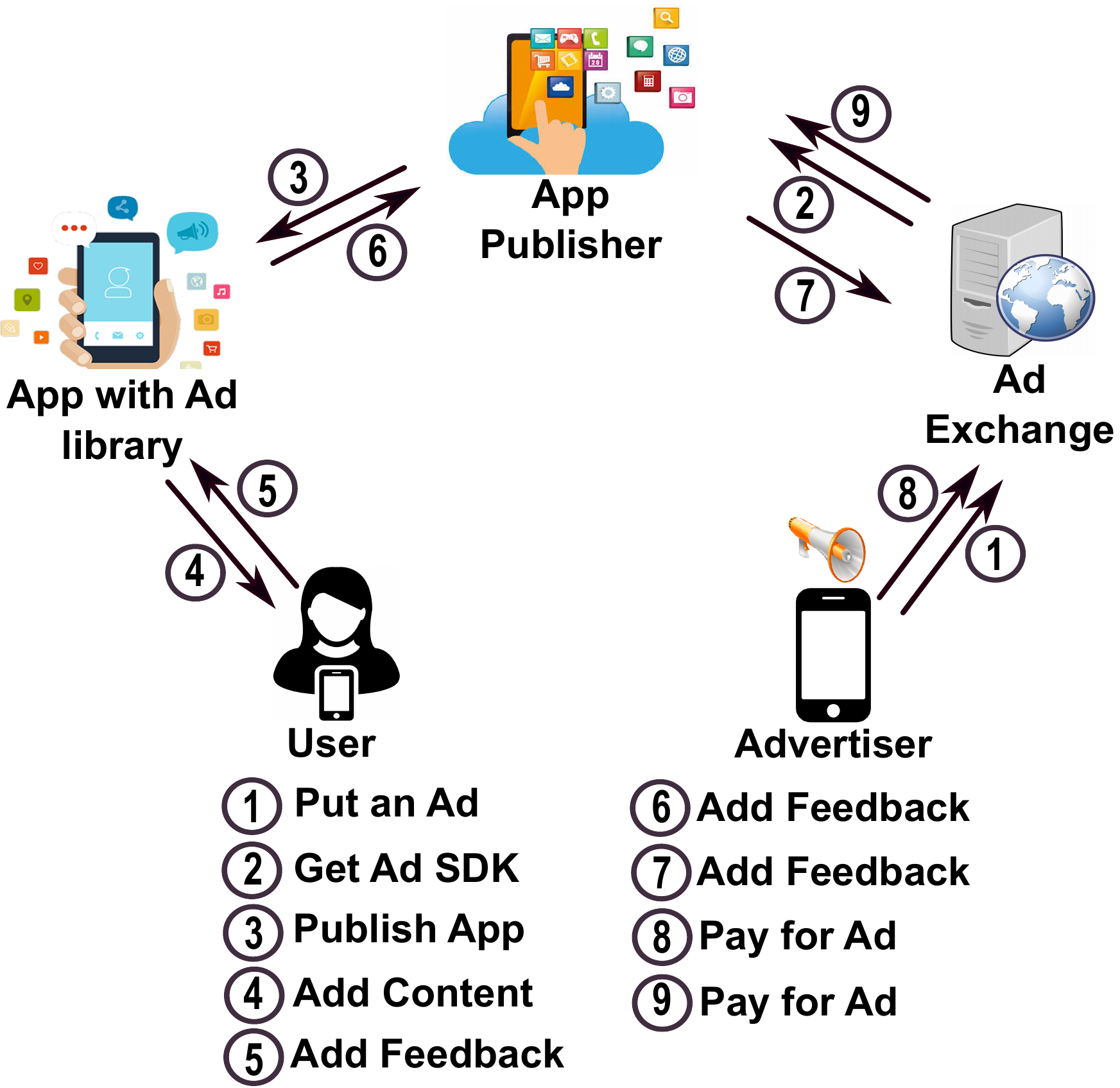}
	\caption{\small The process of serving ads in mobile advertising system.}\label{fig:mobile_serv}
\end{figure}

In the mobile platform, ads are embedded as an advertising library to display ads in applications. The process of ad serving in mobile system is illustrated in Fig.~\ref{fig:mobile_serv}. The advertiser distributes ads to display advertisements to mobile users through the ad exchange (step~1). When a publisher wants to display ads, he must register on the ad exchange to receive the ad library (SDK library) (step~2). The library typically provides an API for embedding ads in the User Interfaces (UI) of the app publisher and fetching, presenting, and tracking ads (step~3). The device ID is usually used to identify the publisher who wants to embed those ads uniquely. Users download the app and run it on a smartphone. Once the program starts, the ad library fetches ad content (step~4) and sends feedback (impression and clicks) to the ad exchange (step~5-7). Advertisers pay advertising networks and app publishers based on the number of impressions or the number of clicks (step~8-9).

\subsection{Targeted Advertising}
\label{sec:targetAd}

The most obvious difference between online advertising and a traditional approach is that the former displays advertisements to the customer based on their interests, while in the latter, advertisements are massively broadcast without considering the customer's interests. Ad networks use ad targeting methods to increase their income, and in this way can display advertisements based on the user's preferences. The three most popular types of ad targeting can be categorized into \emph{contextual}, \emph{behavioral}, and \emph{location-based} approaches.

In the contextual approach, advertisers display relevant advertisements by focusing solely on the content of the web page being viewed by the user~\cite{zhang2012contextual}. A behavioral targeting strategy allows advertisers and publishers to utilize information from the user's browsing history (e.g., by monitoring the behavior of the user on the Internet) to customize the types of advertisements they are served. Whenever an individual visits a website, all of the relevant information, including the pages visited, the period of time spent on each page, the links that are clicked on, and the things that are interacted with, are stored in a profile linked to that visitor~\cite{koran2013behavioral}. Based on the data in these profiles, publishers can show related advertisements to visitors that match their habits. In a location-based targeting, location-specific advertisements are delivered to potential users; this technique is particularly useful for mobile advertising~\cite{chatwin2013overview}. 

Although targeted advertising can be profitable for the advertiser, collecting consumer information raises their privacy concerns. Hence, governments enact more data privacy regulations. For example, the European Parliament has adopted the General Data Protection Regulation (GDPR \footnote[5]{ The GDPR is a provision in the European Union Data Protection and Privacy Act.}) to increase the transparency on how individuals data is used and stored. GDPR is an opportunity to build trust among marketing consumers. Becoming a widespread and trusted brand is critical to maintaining power in today's increasingly competitive world. Using data privacy as a core principle in business enables the business to establish an honest and humane relationship with its customers and partners. On the other hand, stricter regulation can limit advertisers' targeting advertising and slow down progress in the advertising ecosystem. Hence, there must be a trade-off between GDPR rules and targeted advertising to have a healthy dynamic in the advertising ecosystem.

\subsection{Revenue Models}
\label{sec:revenue}

In this subsection, we discuss how entities in the online advertising network generate revenue.

Typically, publishers agree to display an advertiser's advertisements and share the keywords used by the advertiser in their website, charging a commission fee for the action(s) generated by the user. This agreement includes a contract made by a broker (also called an Internet advertising commissioner) between publishers and advertisers. The commissioner also controls the advertisers' budget, to avoid over-spending~\cite{metwally2006hide}. As soon as the advertiser pays the publisher the commission fee, it displays links determined by the advertiser on its website~\cite{mittal2006detecting}.

The general models~\cite{stone2011understanding, chatwin2013overview} used by publishers to make money through advertising are determined based on the numbers of impressions, clicks, and actions. We explain each of these types of revenue model in detail in the following subsections.

\subsubsection{Cost per Impression}
\label{sec:CPI}

This model is favored by publishers and was developed based on traditional advertising systems. A metric called \emph{Cost Per impression Mile} (CPM) is often used to measure the cost per impression, where the advertiser' payment to the ad network is calculated based on the cost of 1,000 views of an ad.

\begin{figure}
	\centering
	\includegraphics[width=0.5\textwidth]{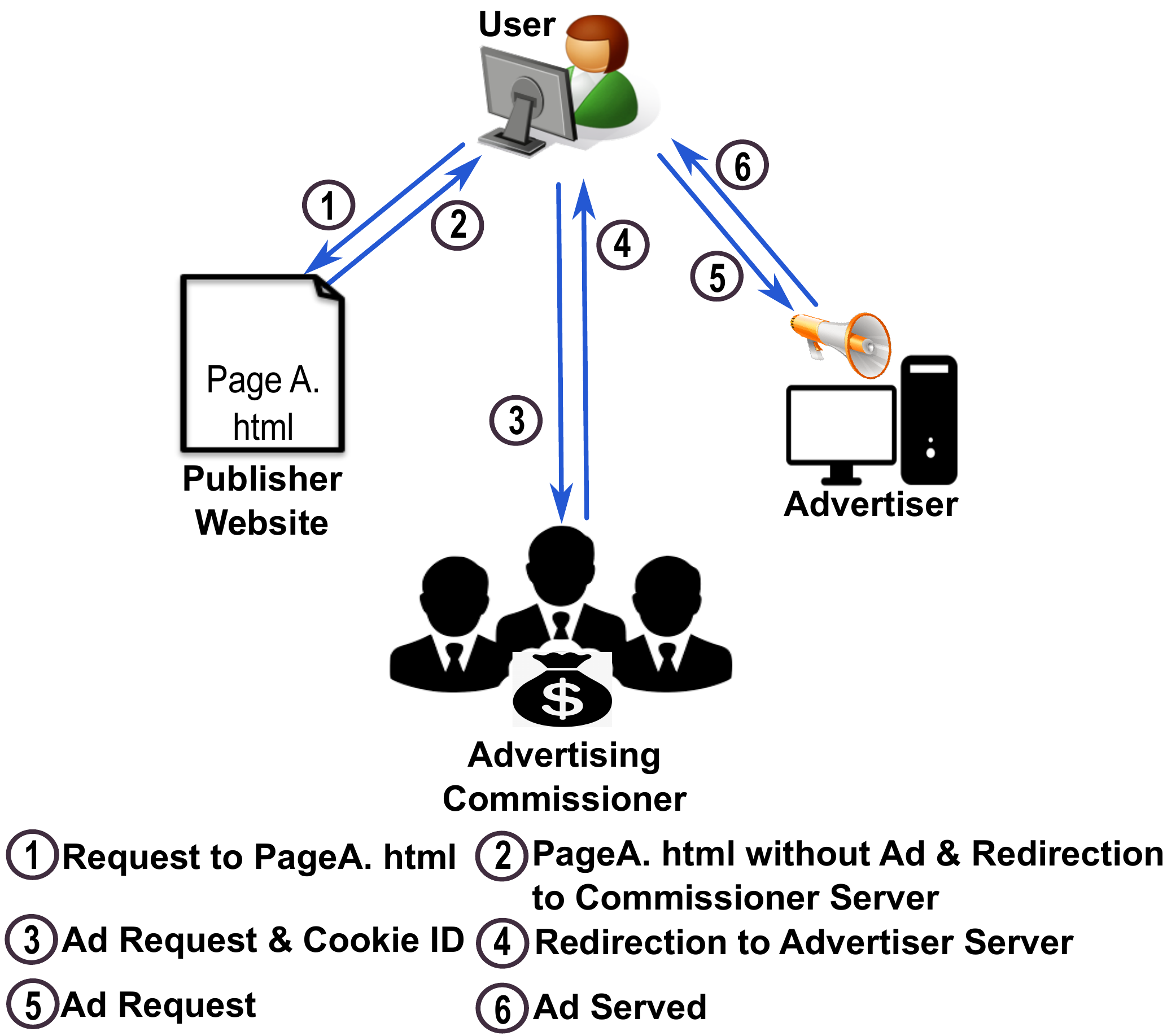}
	\caption{\small Impression traffic model used in online advertising systems.}\label{fig:a_Imp}
\end{figure}

To enable a better understanding of how the commissioners process the receiving impression traffic, Fig.~\ref{fig:a_Imp} illustrates this process. The steps are as follows: (1) a user requests a website; (2) in response, the publisher displays the requested website in the user's web browser; and (3) the user's browser redirects to the commissioner's web server (the commissioner does not repeat advertisements since it stores the recent advertisements shown to the user in browser cookies). In steps~4~and~5, the commissioner allows the user's browser to redirect to the advertiser server. In step~6, the commissioner loads the advertisement into the user's browser.


\subsubsection{Cost per Click}
\label{sec:CPC}

In the Cost Per Click (CPC) model, the advertiser pays the publisher based on how many times a viewer clicks the ad on the publisher's web page. Many search engines, including Yahoo, Microsoft, and Google, prefer to use the CPC model. The reason for that is because a user clicking on an ad is a strong signal of interest; as such, CPC guarantees a better return on investment than CPM, where advertisers pay for their advertisements to be shown without counting on any implicit feedback from users.

The click traffic model is the approach that is most similar to the impression traffic model. However, the process starts when a user clicks on a hyperlink on the publisher's site. Then, the user is redirected to the commissioner's server. The server then logs the click for accounting purposes. After that, the server of the advertising commissioner redirects the user's web browser to the web page related to the advertiser.


\subsubsection{Cost per Action}
\label{sec:CPA}

In general, the CPC charging model is considered to be a specific case of the Cost Per Action (CPA) model, in which the publisher is paid whenever a user-generated click leads to a predefined action being performed, e.g., filling in a form on the page, signing up, registering, or downloading an item corresponding to the ad. Advertisers prefer to deploy this type of cost model since they only pay the publisher for specific actions. Although this approach has advantages for the advertiser, it also has some drawbacks. It is challenging to implement, especially in the case of complex actions, and the publisher is less interested in applying this model since dishonest advertisers may deflate the number of actions to pay a lower commission fee (See Section~\ref{sec:conven_fraud}).

Some software also help advertisers and publishers to generate revenue by these models. One of them is adware (or advertising-supported software), which uses pop-up messages or unclosable windows to display ads. The first group of adware is known as \emph{shareware}. This is designed for consumers who are unwilling to pay for specific software, and numerous ad-supported software, games, and utilities have been distributed as adware. This type of software automatically displays advertisements in the form of annoying pop-up messages, and users have an option to disable these advertisements if they buy a license key. The developer uses the adware to recover the costs of development, maintenance, and upgrading of the software. Also, this approach allows consumers to use the software free of charge or for a low price. The second category can be thought of as a kind of \emph{spyware}~\cite{vratonjic2013security}. This group stealthily collects information on customers by spying on them to serve advertisements embedded in websites. In formal terms, these applications generate revenue for developers by tracking the user's Internet surfing habits to display advertisements associated with the user.

Despite the advantage of adware to make a profit for the developers, it can endanger users' privacy. The adware can collect the required information by continually monitoring the search toolbars of browsers without the user's awareness or permission. In extreme cases, the adware sells this private information to other entities without the awareness or permission of the user. There is a solution to combat adware and enhance the online advertising system's security and privacy, which is called ad-blocking. Ad-blockers are applications that help users passively stop pop-ups ads and banners from displaying in their browsers~\cite{gordon2021inefficiencies}. AdBlock, AdAway, and AdGuard are ad-blocking add-on (or browser extension) software that can be added to the browser to prevent adware.

Early digital advertising efforts developed intrusive formats
such as pop-up ads or autoplaying audio/video ads. This led
to consumer demand for ad blockers, applications that allow
users to passively block advertising from showing up in their
browsers

\subsection{Payment of commissions in online advertising systems}
\label{sec:pay_commi}

In this subsection, we briefly explain how advertisers pay commission fees to commissioners and publishers.

When advertisers receive valid traffic generated from impressions or clicks, they have to pay the publisher. The commissioner also earns a fraction of this income. If the advertiser uses a similar scheme to pay the publisher, then the commissioner's percentage will be calculated at a fixed rate. For example, in the case where an advertiser pays a publisher per click (or impression), and the publisher receives the money based on the number of clicks (or impressions), then the commissioner receives a fixed payment.

However, an advertiser may pay based on the number of sales, while the publisher earns per click (or impression). This practice is known as an \emph{arbitrage campaign}~\cite{metwally2006hide}. In formal terms, an arbitrage campaign is one where the advertiser uses different payment metrics to pay the commissioner and publisher. In an arbitrage campaigns, the commissioner should ensure that its share of the profit from the advertiser is more than the publisher's payment; otherwise, the commissioner loses money. In reality, advertisers prefer to pay based on sales, while publishers prefer to receive income according to the number of impressions or clicks. Hence, Internet advertising schemes are mainly arbitrage campaigns. However, some advertisers may prefer to pay on the basis of clicks or impressions for product branding.

\subsection{Main Security Goals in Online Advertising System}\label{sec:SecurityR}

Cybersecurity aims to protect a company's digital assets against cyber attacks. Cybersecurity can be achieved by using appropriate security controls to provide several security features such as deterrence, prevention and detection of cybercrime. The primary goal of cybersecurity for each system (e.g., online advertising system) is to ensure three principles, including confidentiality, integrity and availability (CIA) of data and services. The CIA is essential in cybersecurity because it provides essential security features, helps avoid compliance, ensures business continuity, and prevents reputational damage to the organization. Confidentiality refers to protect information from unauthorized access. The online advertising system is responsible for protecting consumers' private information. Integration refers to protecting data against deletion or modification by unauthorized individuals or systems to ensure the accuracy and consistency of information. Any tampered data injected into the online advertising system can interrupt the system functionality. The term availability refers to ensuring that the system is available only for permitted users whenever required. Lack of availability in the system can cause severe damages.

\section{Vulnerability of Online Advertising Systems}
\label{sec:attack_adv}


Online advertising systems are vulnerable to various types of attacks, and in this section, we present a taxonomy of current attack methods.

\begin{figure*}
	\centering
	\includegraphics[width=\textwidth]{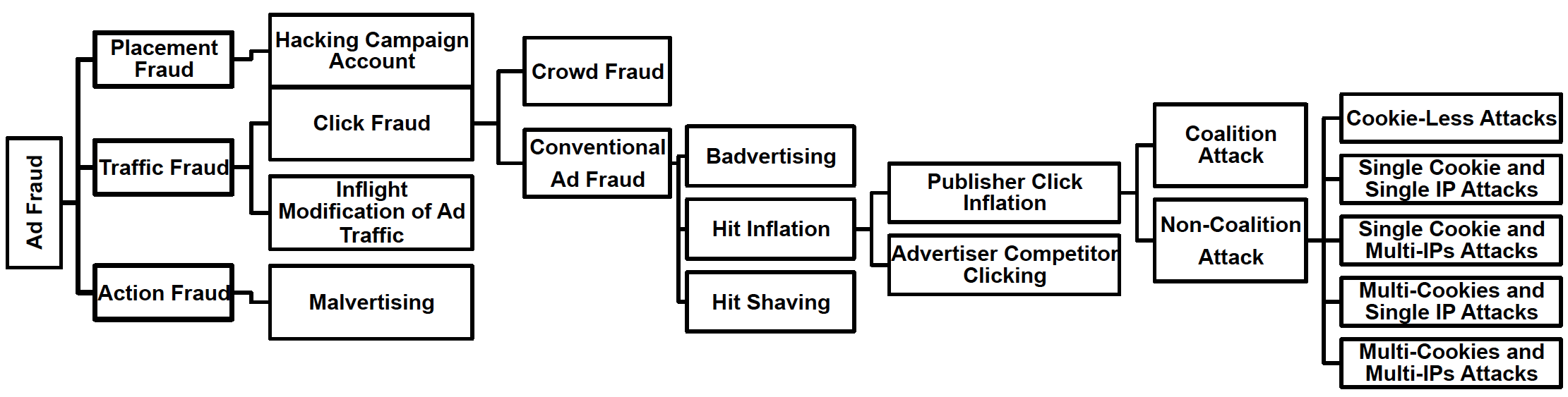}
	\caption{\small Proposed taxonomy of ad fraud attacks in online advertising systems.}\label{fig:taxonomy}
\end{figure*}

Fig.~\ref{fig:taxonomy} illustrates the proposed taxonomy. The taxonomy sets out the major fraudulent activities in response to questions about who does what and how. ``Who'' responds to question about whether the fraud is created directly from human or nonhuman users. ``What'' intends to classify fraud based on the target revenue model. ``How'' responds to how a fraudster performs fraudulent acts for certain types of fraud. Based on the above considerations, we identify three dimensions for classifying advertising fraud. The classification mainly depends on whether the fraud targets ad placements, ad traffic, or user actions. The first type of fraud is placement fraud to manipulate/modify the publisher's websites or the content displayed on users' devices to increase impressions or clicks~\cite{zhu2017fraud}. The second type, called traffic fraud, aims to create fake traffic to increase the number of impressions or clicks generated from separate sites or locations. For example, fraudsters can increase impressions and clicks on publishers' websites by using botnet or crowd. The third type is called action fraud that targets users' actions in order to generate revenue. For example, attackers may hire people to download or send forms to convert or send fake cookies to receive commissions as affiliates using robots. We classify online advertising attack methods into four main categories: hacking campaign account~\cite{frik2017economics}, click fraud, inflight modification of ad traffic, and malvertising.
 

\begin{figure}
	\centering
	\includegraphics[width=0.45\textwidth]{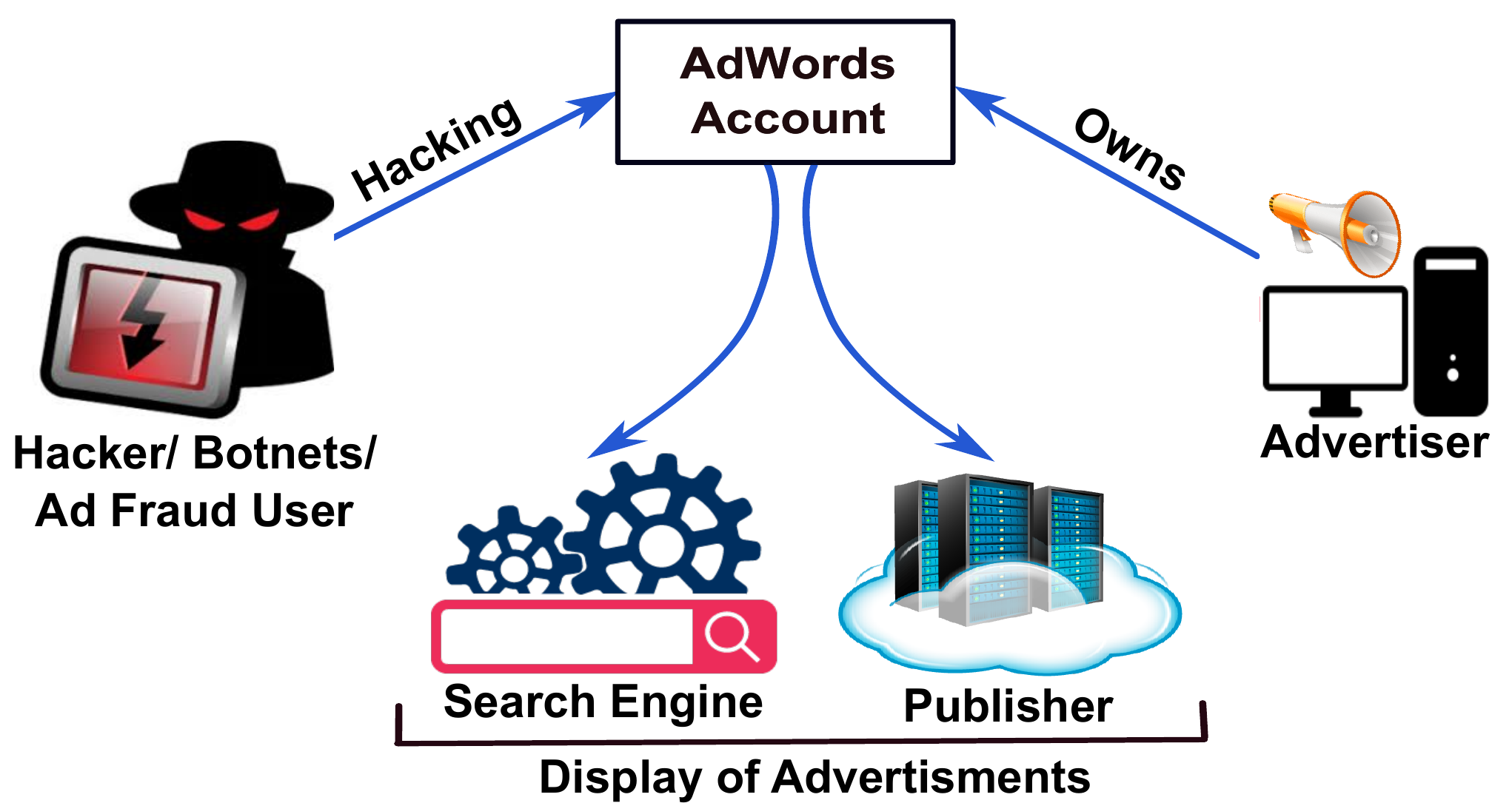}
	\caption{\small Hacking campaign account attack in online advertising network.}\label{fig:hacking}
\end{figure}

\subsection{Hacking Campaign Account}
\label{sec:hack}

The threat of hacking in online advertising arises due to unauthorized access to campaign accounts~\cite{mladenow2015online}. In search engine advertising, companies aim to attract customers by improving the visibility of their advertisements in results pages. Online campaigns can quickly adapt information in their ad campaigns, which are more flexible, targeted, and tailored than traditional marketing campaigns. The flexibility and time savings of online campaigns guarantee that the transaction processing will be fast. An example of this is AdWords~\cite{AdWords}, a tool developed by Google to allow advertisers to create online campaigns in only a few minutes. However, despite all the above advantages to online business, online campaigns face with many challenges, including security and privacy.


We illustrate this with an example. Consider the case where an advertiser creates an AdWords account. Users navigate via the web to run search queries, and advertisements can be presented on the websites of publishers or on the search engine network. If an adversary takes control of the advertiser's AdWords account to launch an attack, this is known as hacking campaign account. The consequences of campaign accounts being hacked include blocking, limited access or unauthorized entry to the account of the advertiser. The availability of short-term online campaigns will also be limited. These results may lead to significant reputational damage, loss of money, and violations of user privacy. Fig.~\ref{fig:hacking} illustrates the hacking of an advertiser's AdWords account.

\subsection{Click Fraud}
\label{sec:click_fraud}

Online advertisements help to develop a healthy Internet, since they provide financial support for the online businesses. The emergence of \emph{click fraud} (also known as malicious clicks, or click spam~\cite{li2014search}) therefore poses a serious security risk to the Internet ecosystem. Click fraud refers to cybercrime activity that is carried out either manually (using human clickers) or automatically (software-supported) to generate fraudulent clicks on the advertisement to make illegal profits.

Fraudulent clicks can damage the health of online businesses, since these clickers can increase their profits or deplete the advertising budgets of their competitors. They achieve this by clicking on advertisements with no actual interest in the content.

In the manual approach, fraud consists of hiring a group of people to increase fraudulent traffic, while automatic click fraud attack is usually based on the use of \emph{botnets\footnote[6]{ To build a botnet, a botmaster (an entity that controls the botnet remotely) needs a network of software robots -- i.e., bots -- that are run independently and automatically.}}~\cite{rodriguez2013survey}. Malicious software called a ``\emph{clickbot}''~\cite{daswani2007anatomy} is one example of this use of botnets to generate fraudulent clicks automatically~\cite{kantardzic2008improving},~\cite{alauthman2020efficient}. Using a clickbot to launch a click fraud attack is more efficient than the manual type of attack, since it can perform automatic clicking over a time period of several minutes to avoid detection. We categorize click fraud into two types, \emph{crowd fraud} and \emph{conventional ad fraud}, as described the following subsections.

\subsubsection{Crowd Fraud}
\label{sec:crowd_fraud}

The emergence of \emph{crowdsourcing}~\cite{kamar2012combining} has led to a novel form of fraud in online advertising, since it can broadcast a large number of tasks to a numerous online workers. Due to the openness of crowdsourcing systems~\cite{choi2016detecting, doan2011crowdsourcing, tahmasebian2020crowdsourcing}, a crowd of workers can easily be recruited via malicious crowdsourcing platforms to perform an attack against a competitor or to increase their advertising expenses. There are many differences between automatic fraudulent behaviors (conventional fraud), and frauds carried out by humans. For example, a vast number of workers via crowdsourcing platforms can be involved in human-generated fraud, while automatic fraudulent traffic can be deployed relatively few machines. A difficulty also arises in differentiating normal and no distinct traffic induced by real humans from the noisy traffic generated by machines. Methods used to detect conventional fraud therefore fail to identify these human-generated frauds. The phenomenon of exploiting a group of real humans to increase fraudulent traffic in online advertising is termed \emph{crowd fraud}~\cite{tian2015crowd}.

\subsubsection{Conventional Ad Fraud}
\label{sec:conven_fraud}

In contrast to crowd fraud, which is carried out by large numbers of attacking machines, normal and no distinct click behaviors by each web worker, the limited fake traffic generated by each web worker, conventional forms of advertising fraud often have specific features in terms of individual behavior patterns, with few sources and large amounts of traffic. In this regard, the detection of conventional fraud is more straightforward than crowd fraud~\cite{tian2015crowd}. 

We divided the conventional advertising frauds shown in Fig.~\ref{fig:taxonomy} into three categories: \emph{badvertising}, \emph{hit shaving}, and \emph{hit inflation}. A brief overview of how these attacks are carried out on online advertising ecosystems is given below. 

\begin{itemize}[leftmargin=*]
	\item \textbf{Badvertising.} Gandhi et al. defined badvertisement as a kind of camouflaged click fraud attack on the advertising industry~\cite{gandhi2006badvertisements} that silently and automatically generates click-through on an advertisement when users visit the website. This attack can not only remain undetected by web publishers, but also does not compromise the user's machine. Unlike a traditional malware-based click fraud attack~\cite{miller2011s}, badvertisement is a stealthy offense in the form of a malicious mutation of spam and phishing~\cite{jagatic2007social} attacks, except that this attack targets the unaware advertiser as the victim rather than an individual. This is very worrying, since it is easier for an attacker to deceive an individual into visiting a web page than to damage a machine with malware.

	This attack artificially and stealthily increases the number of clicks on ad banners hosted by the fraudster or unaware associates to generate more revenue for the attacker through advertising. The revenue generated in this way is transferred from the advertiser to the hosting websites by the fraudster.
	
	Badvertisement has two main components: $(i)$ delivery, which either transfers consumers to corrupt data or corrupt data to consumers; and $(ii)$ execution, which automatically and invisibly displays advertisements to a targeted user. This stealth attack can be accomplished by corrupting the JavaScript code that is downloaded and executed by the client's browser to publish sponsored advertisements~\cite{zhang2008detecting}. Online advertisement systems typically work by placing a JavaScript snippet file into a publisher's web page. Whenever a user visits this page and downloads an advertisement from the ad server, the JavaScript file will be executed. Downloading the ad causes the frame in the JavaScript file to be rewritten with the HTML code required to show the advertisement. The publisher relies on the click-through payment process to count the number of times the user clicks on the link to the ad provider's server. Finally, the user is referred to the ad client's website. This scenario is illustrated in Fig.~\ref{fig:Normal_Ad}.
	
		\begin{figure}
			\centering
			\includegraphics[width=0.3\textwidth]{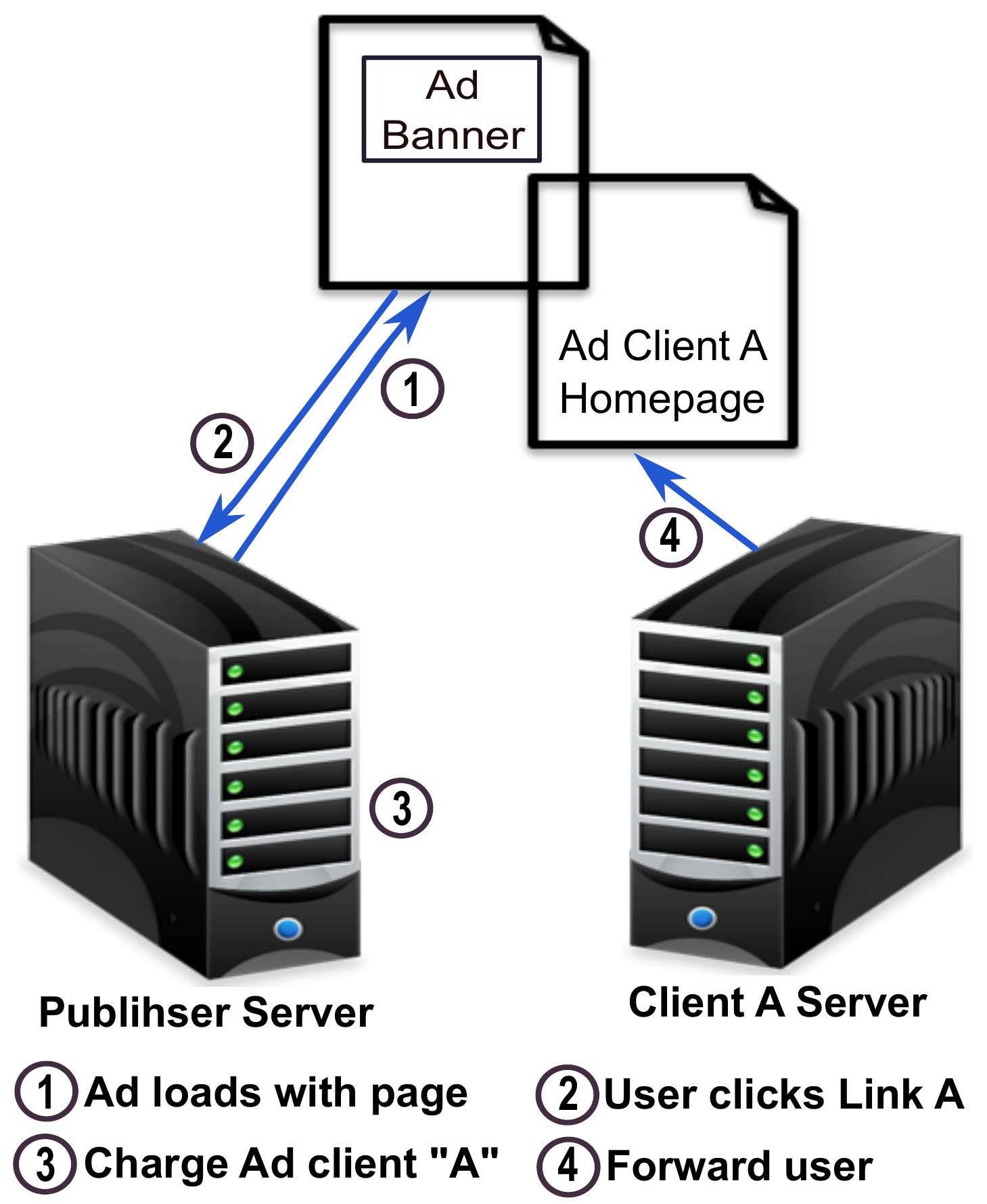}
			\caption{\small Typical online advertisement services.}\label{fig:Normal_Ad}
		\end{figure}

Badvertisements run extra malicious scripts to automatically deploy clicks. In a nutshell, after running the script code and rewriting the frame, the malicious script parses the HTML code and compiles all links. It then changes the web page to embed an HTML \emph{iframe}. If the user decides to click the link, the \emph{iframe} will be activated in the background, and loads its content to exploit the user (Fig.~\ref{fig:hidden_Badvert}).

	\begin{figure}
		\centering
		\includegraphics[width=0.3\textwidth]{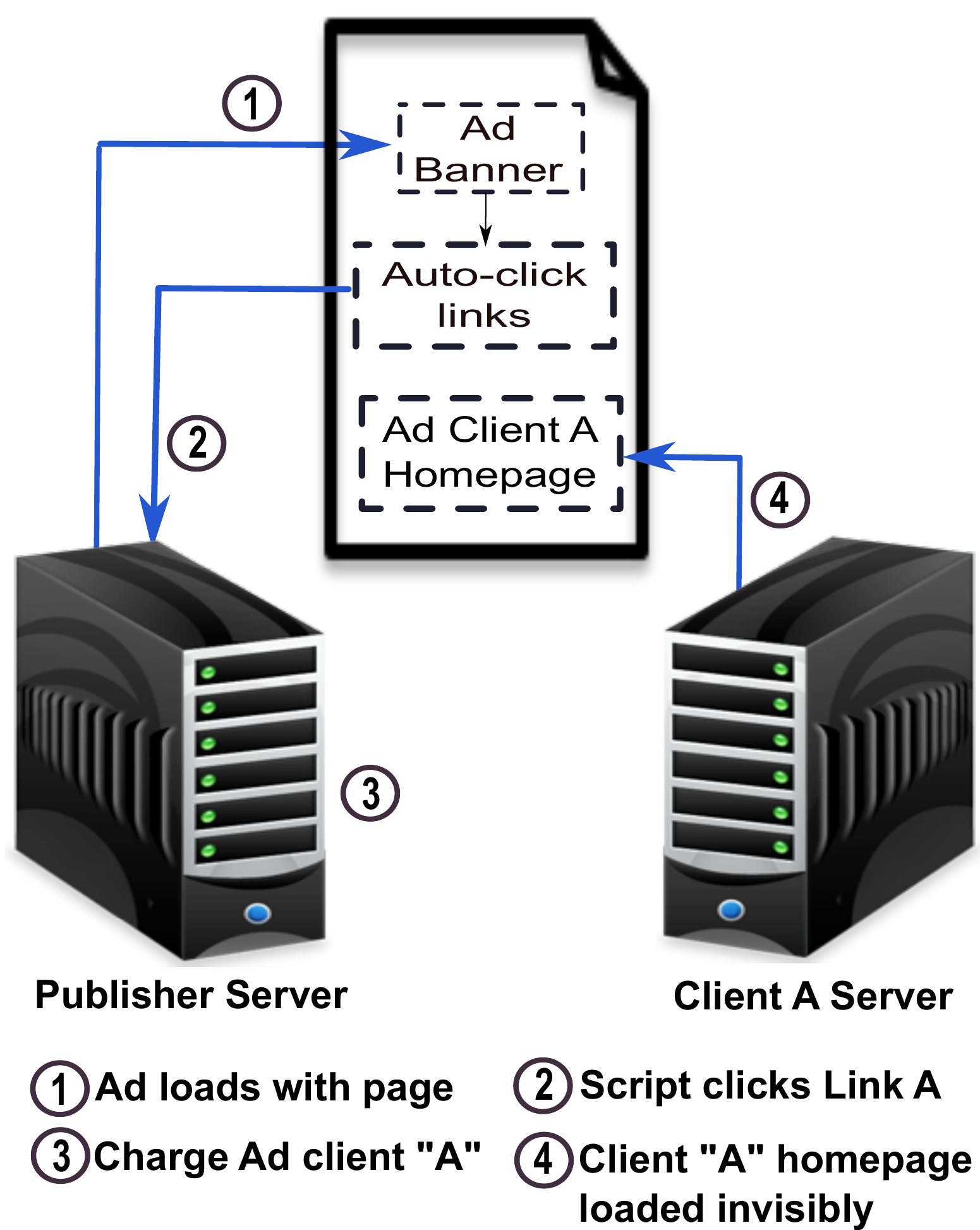}
		\caption{\small Auto-clicking in a hidden badvertisement.}\label{fig:hidden_Badvert}
	\end{figure}

\item \textbf{Hit Shaving.} As previously mentioned, advertisers often prefer the CPA model to pay the publisher based on the desired user action, rather than for each click on their ad. However, the CPA model is vulnerable to hit shaving (also called \emph{deflation fraud}~\cite{ding2010hybrid}). In this attack, a fraudulent advertiser undercounts the real transactions to pay a lower commission fee. 
	
Before describing how the hit shaving attack is applied in an advertising network, we need to give an overview of the mechanisms used in click-through payment programs.

Advertising has become a pivotal technology on the Internet, as confirmed by the growth of click-through payments. The main entities involved in click-through payment programs are the user who views the page and clicks on a link, the referrer who exposes advertising material to the user, and the target site running the click-through payment process.

A click-through payment system works as: we suppose that there are two websites A and B, and that A can refer the user to B. Hence, whenever B receives a referral from A, B has to pay the webmaster\footnote[7]{ The webmaster is the person controlling the content served to the user.} of A for this reference. In more detail, when a user views web page A and clicks on a link that refers the user to web page B, then A should receive money from B. In other words, the user has ``clicked-through'' A to reach B. The use of a click-through payment program by the webmaster of B leads to an increase in traffic to the website, since other websites display links to B. However, since the underlying infrastructure of this structure is based on the HTTP protocol, it is exposed to attack.

For a better understanding of how this mechanism is vulnerable to fraud, we review the procedure used to exchange HTTP messages (see Fig.~\ref{fig:HTTP}) during a click-through event. As illustrated in Fig.~\ref{fig:HTTP}, when users view a web page from site A (called the referrer), the HTTP procedure is executed. Site A includes a link to site B (called the target), and agrees to take part in the process of click-through payment to site B. The customer's browser sends a request to load the page from site B when the link is clicked. Site B can identify the site from which the requested web page originated (i.e. where the user are is being referred from) simply by checking the referrer field in the HTTP header.

The previous explanation should reveal that the click-through payment system has the potential to be exploited for fraud. The problem arises from the lack of communication between A and B after the user clicks on the link. A cannot verify how many times its web page has referred users to the targeted page, and as a consequence, B is able to omit some of the click-through events from the referrer, in a scheme called \emph{hit shaving}. In addition, although the referrer site can detect that the target site has shaved its referrals, it cannot provide proof of this to a third party. A can also conduct fraud against B by generating false requests in order to increase the payment from B, and this is called \emph{hit inflation}. In brief, hit shaving is a form of fraud by a dishonest advertiser who can undetectably change the number of clicks received from a publisher in order to pay a lower commission fee~\cite{metwally2007hit, metwally2006hide}. 
	
	\begin{figure}
		\centering
		\includegraphics[width=0.35\textwidth]{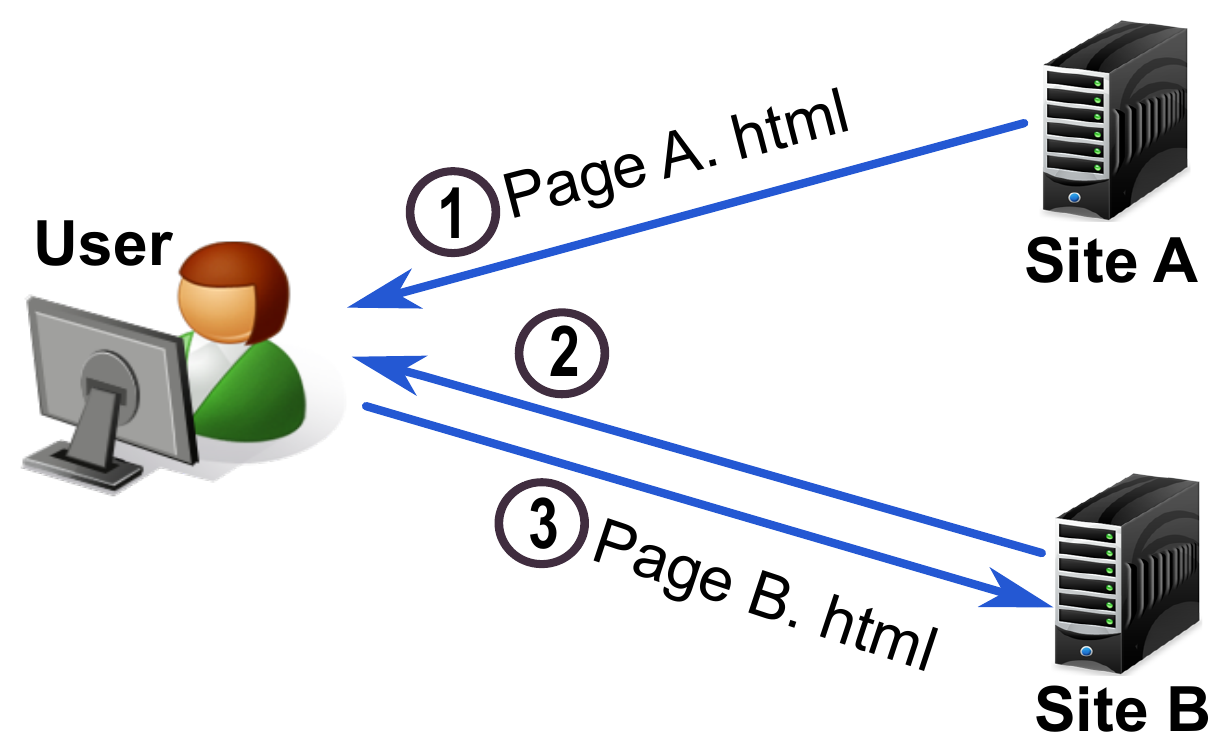}
		\caption{\small Workflow for a click-through system. Step~1: user retrieves Page~A.~html from site A (referrer site). Step~2: user clicks on a link in site A and requests the page from site B (target site). Step~3: Page~B.~html on site B will be uploaded for the user.}\label{fig:HTTP}
	\end{figure}

	\item \textbf{Hit Inflation.} This is a fraudulent activity performed by an adversary to inflate the hit count, in order to boost revenue or hurt competitors.

	In~\cite{anupam1999security}, a sophisticated type of hit inflation attack is defined that is very hard to detect. Fig.~\ref{fig:hit_inf} illustrates this attack scenario, which involves an association between a fraudulent website (W) and a fraudulent publisher (P), where W uses a script code to silently divert a user to P. The scenario starts when a user simulates a request or click to fetch page~W.~html from W (step~1). However, the user is redirected to page~P.~html (step~2). P has two forms of the web page: a manipulated form and a valid form. P will show a manipulated web page to the user when the referrer field in the HTTP request shows W (step~3) and clicks the ad by itself without knowing the user. Otherwise, P will direct the user to the valid web page, and the user is free to either click on the ad or not (step~4,~5).
	
	Publishers and advertisers are the two entities in online advertising systems that are the major sources of inflation attacks. The two most common types of hit inflation attack are called \emph{publisher click inflation} and \emph{advertiser competitor clicking}. We briefly illustrate both types of attack below.
	 
		\begin{figure}
			\centering
			\includegraphics[width=0.4\textwidth]{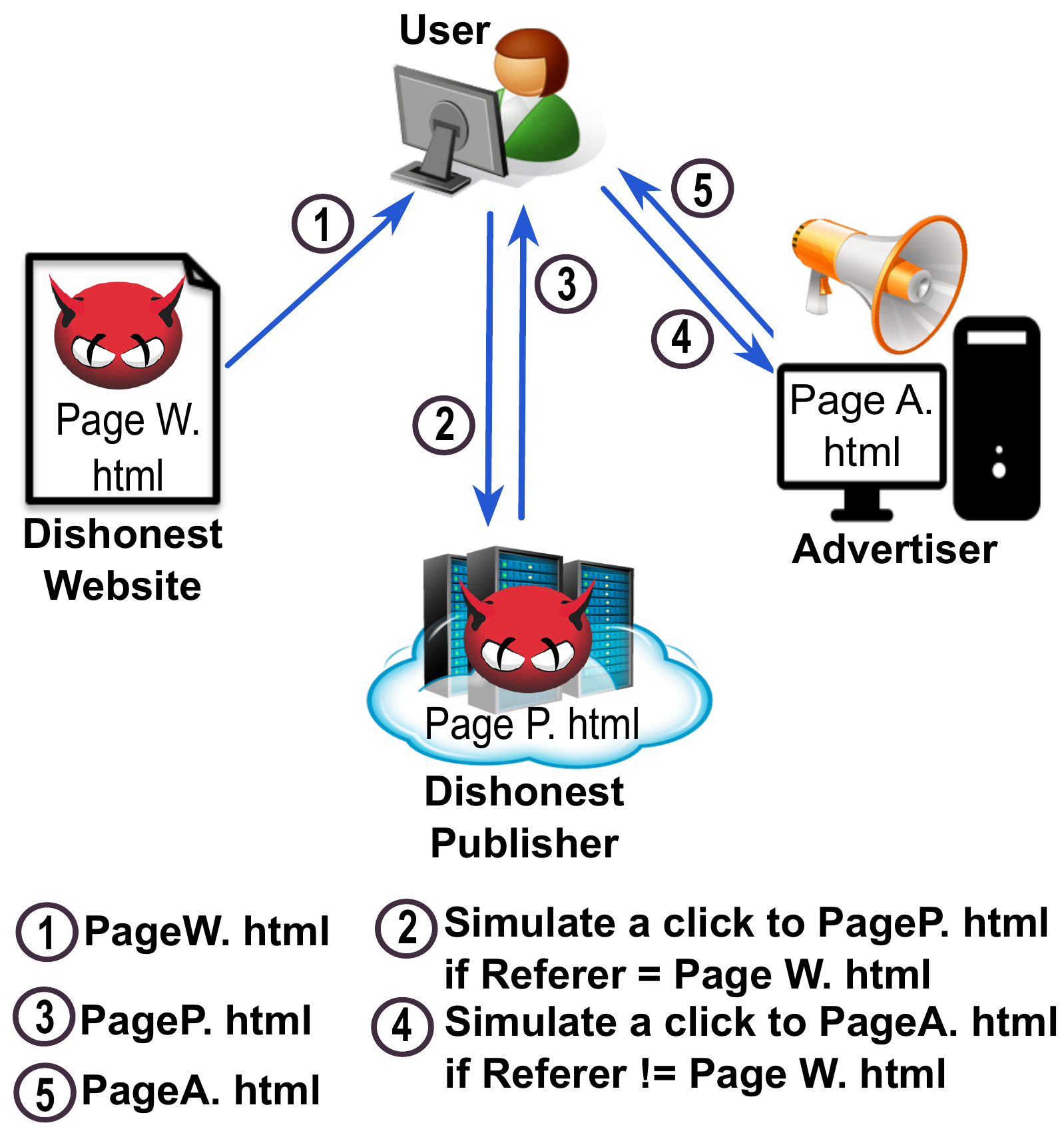}
			\caption{\small Hit inflation attack on online advertising network.}\label{fig:hit_inf}
		\end{figure}
	
		\textbf{\emph{Publisher click inflation}}. In publisher click inflation, a dishonest publisher is motivated to artificially inflate the click-through count (without real interest in the content of the advertisement) to obtain more income from ad networks. As discussed earlier, if the advertiser wants to present its advertisements on the publisher's website, the publisher enters into a contract with the broker (commissioner). The publisher then gains income from advertisers through the user-generated traffic that they send to websites of advertisers. Obviously, the more clicks the publishers earn, the more money they generate. Consequently, this opens the way for malicious publishers to create illegal revenue by increasing the numbers of clicks, impressions, and actions on their websites.

Publisher click inflation attacks can be classified into two categories: \emph{non-coalition} and \emph{coalition} attacks~\cite{oger2015privacy}. The former is performed by a single publisher (one fraudster) who solely generates traffic to its resource(s), while the latter involves a coalition attack among a group of publishers who share their systems. If we can detect both categories of attack, we can claim that the problem of hit inflation is solved.


Launching a coalition attack has several benefits for fraudsters. Firstly, the possibility of fraud detection decreases because the attackers do not need to reuse their resources to generate more attacks~\cite{kim2011catch}, making it difficult for detection algorithms to identify the relationships (e.g., the relationships between the cookie IDs and IP addresses of the resources generating traffic and the sites of fraudsters) between each fraudster and all the attacking machines. Secondly, the cost of launching an attack is reduced by sharing resources rather than increasing the number of physical resources.
		

The study in~\cite{metwally2006hide} classifies non-coalition attacks according to the number of IPs and the cookie IDs of the system, and the way in which the commissioners recognize the machines of the surfers (potential Internet customers). When customers visit a website, this traffic has certain fixed characteristics which are different from automatic traffic, and typically involve relationships between IP addresses and cookie IDs. Hence, if fraud detectives find inconsistencies between the cookie IDs and the IP addresses, they can investigate manually by selecting a subgroup of the publishers to detect the attack. On the other hand, when dishonest publishers want to launch the attack, they can leave a false fingerprint for the relationship between the IPs and cookie IDs in order to confuse the detection mechanisms.

The attack can be launched by one or multiple IPs, and these addresses may be associated with no, one or multiple cookie IDs. There are therefore six possible types of attack based on combinations of IPs and cookie IDs, as follows.

		\begin{enumerate}[leftmargin=*]
			\item \textbf{Cookie-Less Attacks.} A fraudster can launch cookie-less attacks in at least two known ways. Firstly, there is the option for the attacker to turn off cookies on the system(s) which plan to launch the attack. Secondly, a fraudster can employ commercial services called network anonymization, which are designed to protect the privacy of users~\cite{broder1999data} and to block third party cookies to give more cookie-less traffic.
			
			\item \textbf{Single Cookie and Single IP Address Attacks.} In this type of attack, a dishonest publisher can employ a script to launch an attack from one machine with a fixed IP and one cookie ID. The author in~\cite{klein1999defending} provided an example of this type of script.
			
			\item \textbf{Single Cookie and Multiple IP Addresses Attacks.} Attacks of this type are more widespread among fraudulent advertisers than fraudulent publishers, since changing the IP address of the attacking machines is more convenient than changing the cookie ID. The commissioner shows the most profitable advertisements to Internet customers that have not recently been displayed. In addition, if repeating the same cookie sends to the commissioner, as a consequence, the same advertisements display to the users. Hence, a dishonest advertiser can start the attack by visiting the publisher's website and continuing until the broker shows advertisements from its competitors. The fraudster then stores the cookie ID with the intention to continuously applying the ID to force the broker to show the advertisements from its competitors. In this way, it can simulate clicks on advertisements in order to drain its competitors' advertising budgets.
			
			\item \textbf{Multiple Cookies and Single IP Address Attacks.} An attacker can perform this type of attack in various forms. The simplest method is to connect different systems to the Internet via a single router, and then execute various scripts on the systems. In this way, the attacker can simulate receiving traffic with several cookie IDs but a single IP address. However, this type of attack is not economically viable. This attack suffers from a resemblance to the regular Internet traffic problem, in which different customers connect to the Internet with various cookie IDs using a single IP address through an Internet Service Provider (ISP).

			In the second form, in order to make the attack more comprehensive and sophisticated, the attacker can connect several machines to the Internet via an ISP with a similar IP. To reduce the impact of this malicious attack and defraud the detection algorithms, a dishonest publisher can combine fraudulent traffic with regular traffic.
			
			\item \textbf{Multiple Cookies and Multiple IP Addresses Attacks.} Performing and detecting this class of attack is difficult. The malicious publisher uses various valid cookies and IPs. The attacker can perform this type of attack by using the cookies and IPs in multiple forms. In the most simple form, which is not economically viable, the attacking publisher has access to various machines with different accounts with ISPs. Another method is to use botnets, such as spyware and Trojans. The aim of using a botnet~\cite{shaw2003spyware} is to simulate impressions and clicks on the website of the attacker by sending the proper HTTP requests while exploiting the cookies and IPs of legal users. The traffic generated in this way is very similar to regular traffic.

This type of attack can be considered a more sophisticated version of some of the above examples. Suppose that the publisher has access to different legal cookies and IPs, such that IPs can generate random or can be pre-assigned. Then, whenever a cookie ID and a pre-assigned IP is used in the attack, the attack can be considered a more sophisticated version of the multiple cookies/single IP attack that uses multiple IPs. In contrast, when the IP is selected randomly, this results in the use of identical cookies for different IPs. This attack can also be considered a more sophisticated version of the single cookie/multiple IPs attack with multiple cookies.
		
		\end{enumerate}
		
		\textbf{\emph{Advertiser competitor clicking.}} In this attack, malicious advertisers carry out hit inflation attacks against their competitors to drain their advertising budgets. In the case where competitors have limitations on their daily advertising budget to participate in bidding, fraudsters can increase the probability of their advertisements being displayed by winning the auction.

\end{itemize}

More generally, the consequences of fraudulent traffic include reducing the reputation of the commissioner and attracting fewer advertisers, and also may lead to extra fees or penalty payments for advertisers~\cite{johnston1976cliques, kannan2004clusterings}.

\subsection{Inflight Modification of Ad Traffic}
\label{sec:Modi_Ad_fraud}

In~\cite{vratonjic2011online}, a new form of ad fraud was presented that involves the inflight modification of advertising traffic (also called a Man-In-The-Middle (MITM) attack). An well-known example of this type of fraud is the Bahama botnet, which allows malware to force compromised machines to show surfers altered advertisements, and to change the results of searches~\cite{Botnet}. The key difference between this attack and traditional click fraud is that in the latter case, ad networks can gain income from fraudulent clicks, while inflight modification of ad traffic can allow either traffic or income to be diverted from the ad networks to the attacker's server.

In the Bahama botnet, compromised systems direct users to a malicious site that looks identical to real Google search results. In this case, the attacker leads the user traffic to another site of the attacker's choosing, such as a fake website, by corrupting the translation of the Domain Name System (DNS) on the infected systems. For example, when a compromised user clicks on advertisements on Yahoo or Google, they are silently redirected to a server that is under the attacker's control. Consequently, the domain name/hostname ``Google.com'' (or Yahoo.com) translates to an IP address that belongs to the attacker and not to Google (or Yahoo).

Moreover, a viewer can enter a query into the input box that appears to belong to the Google server, but the traffic is in fact redirected to the poisoned server. The user is sent back (malicious) results for the given query from Google, i.e. results that are different from the real ones. Clicking on these fake results leads to the click-through payment program being triggered, and thus to advertisers receiving money, meaning that click fraud has taken place. In the case of Bahama botnet, income is diverted from main ad networks to smaller publishers and ad networks.

The adversary can also use botnets of compromised wireless routers rather than compromising the users' systems~\cite{Botnet2}. In this scheme, the wireless router, which is hacked by malware, is converted to a bot. The botnet master can then give instructions to launch an inflight modification of traffic attack to transmit traffic through the router. Many public hotspots operate on this model by providing users with free Wi-Fi while embedding advertisements in the users' traffic to earn more money.

Inflight modification of ad traffic has a drawback in that if a user clicks on the displayed advertisements, profit is generated for the fraudster rather than the legal ad network. Hence, this attack weakens the network industry model. It is worth noting that there are other catastrophic effects of these attacks in terms of the security of end-users (as it leads to malvertisement rather than legitimate advertisements), and also a loss of reputation and income for legal advertisers.

\subsection{Malvertising}
\label{sec:malver}

The primary goal of the online advertising system is to reach users, and these entities are therefore more vulnerable to threat in this system than the others. 



When a user navigates the Internet and visits different websites associated with a single advertiser, the same cookies are allocated to the user. In this way, the ad provider can track the user's online activities by compiling the information from the cookies without the user's permission or consent. The consequence of this tracking is that the user's privacy is violated. Moreover, users can be involved in fraud (e.g., click fraud) without realizing. Malvertising (malicious advertisements) is another fast-growing security threat on the web that can infect users~\cite{vratonjic2011online}. 

Malvertisement is a platform for distributing malware by injecting malicious code into legitimate ad networks. Malware can be categorized as worms, viruses, Trojans, rootkits, ransomware, botnet, etc~\cite{qamar2019mobile}. Malicious ads take advantage of browser vulnerabilities to infect the victim's system, persuade users to download and to install malicious software or redirects users to websites they have not planned to visit (in the case of ransomware)~\cite{bermudez2020measurement}.


As previously mentioned, there are several entities involved in an online advertising system, making it a complex network. This complexity and the use of multiple redirections between different components allows attackers to embed malicious content (e.g., malicious advertisements) in places that publishers and ad networks would not expect. For example, an report by Blue Coat~\cite{larsen2010exploiting} shows that JavaScript code can be served by an ad server to inject a hidden iframe tag into a benign site instead of fetching legitimate advertisements. In this scheme, the iframe commands the browser of the victim to silently interact with a malware server, allowing a PDF exploit file to be downloaded. Both publishers and advertisers in the online advertising ecosystem have the potential to launch a malvertising threat; for instance, an advertiser can easily inject a malicious ad into a legal ad network to trigger malvertising. As a result, the advertising network may deploy those advertisements on publishers' websites, and users will then access them by clicking. Moreover, publishers can insert malicious content into their sites to indirectly cause a consumer to install malware. In this scheme, users even do not need to click on advertisements to activate malware. One of the most common forms of malvertising is flash-based advertisements~\cite{ford2009analyzing}, in which an Adobe Flash File (also referred to as a SWF) that contains malicious script is abused by criminals to run arbitrary commands. Creating advertisements with animation and sound in an SWF file allows the advertisers to attract a greater audience, and this means that Flash is vulnerable to being used in malicious attacks. It is therefore clear that attackers can spread malicious advertising via Flash, which is known as ``malvertising''~\cite{ford2009analyzing}.

\begin{table*}[]
	\caption{\small Summary of attacks, description, attack goal, revenue model goal and primary component targets in online advertising system.}
	\label{table:tbl_attack}
	\centering
	\scriptsize{
		\setlength\tabcolsep{1.3pt} 
		\begin{tabular}{|p{2.3cm}|p{4.3cm}|p{4.3cm}|c|c|c|c|c|c|c|}
			\hline
			& \multicolumn{1}{c|}{} & \multicolumn{1}{c|}{} & \multicolumn{3}{c|}{\textbf{Revenue Model Goal}} & \multicolumn{4}{c|}{\textbf{Primary Component Targets}} \\ \cline{4-10} 
			\multirow{-2}{*}{\textbf{Attack}} & \multicolumn{1}{c|}{\multirow{-2}{*}{\textbf{Description}}} & \multicolumn{1}{c|}{\multirow{-2}{*}{\textbf{Attack Goal}}} & \textbf{CPC} & \textbf{CPM} & \textbf{CPA} & \textbf{Advertiser} & \textbf{Publisher} & \textbf{User} & \textbf{Ad Network} \\ \hline
			\rowcolor[HTML]{EFEFEF} 
			Hacking Campaign Account~\cite{mladenow2015online} & Unauthorized access to campaign accounts&Hacker aims at taking over control of advertiser's account& \xmark & \xmark & \xmark &  \cmark & \xmark & \xmark & \xmark \\ \hline
			
			Crowd Fraud~\cite{tian2015crowd} & Malicious behaviors by humans against competitors for specific targets& Increase fraudulent traffic & \cmark & \cmark & \cmark & \cmark & \cmark & \xmark & \xmark \\ \hline
			\rowcolor[HTML]{EFEFEF} Badvertising~\cite{gandhi2006badvertisements} & 
			Utilizing malicious JavaScript code to publish invisible automatic advertisements in the user's browser&Increase the number of clicks& \cmark & \xmark & \xmark & \cmark & \xmark & \xmark & \xmark \\ \hline
			
			Hit Shaving~\cite{reiter1998detecting} & Dishonest advertisers claim that they received less traffic than in reality & Dishonest advertisers omit to pay commission on some of the received traffic to the publisher & \cmark & \xmark & \cmark& \xmark & \cmark & \xmark & \xmark \\ \hline
			
			\rowcolor[HTML]{EFEFEF} Hit Inflation~\cite{metwally2007detectives} & Artificial inflation of the actual amount of traffic & Economic advantage from over-counting the numbers of transactions & \cmark & \cmark & \cmark & \cmark & \cmark & \xmark & \xmark \\ \hline
			\rowcolor[HTML]{EFEFEF}
			Inflight Modification of Ad Traffic~\cite{vratonjic2011online}& Infecting the system to show altered search results along with modified advertisements to the users& Generate revenue fraudulently for ad networks and publishers& \cmark & \xmark & \xmark & \cmark & \cmark & \cmark & \cmark \\ \hline
			
				Malvertising~\cite{vratonjic2011online} & Perpetrators inject malicious code into legitimate online advertising networks  to spread malware & Malicious code, eventually, attempts to redirect
				users to malicious websites & \xmark & \xmark & \xmark & \xmark & \xmark & \cmark & \xmark \\ \hline
		\end{tabular}}
	\end{table*}
	
	\begin{figure*}
		\centering
		\includegraphics[width=0.6\textwidth]{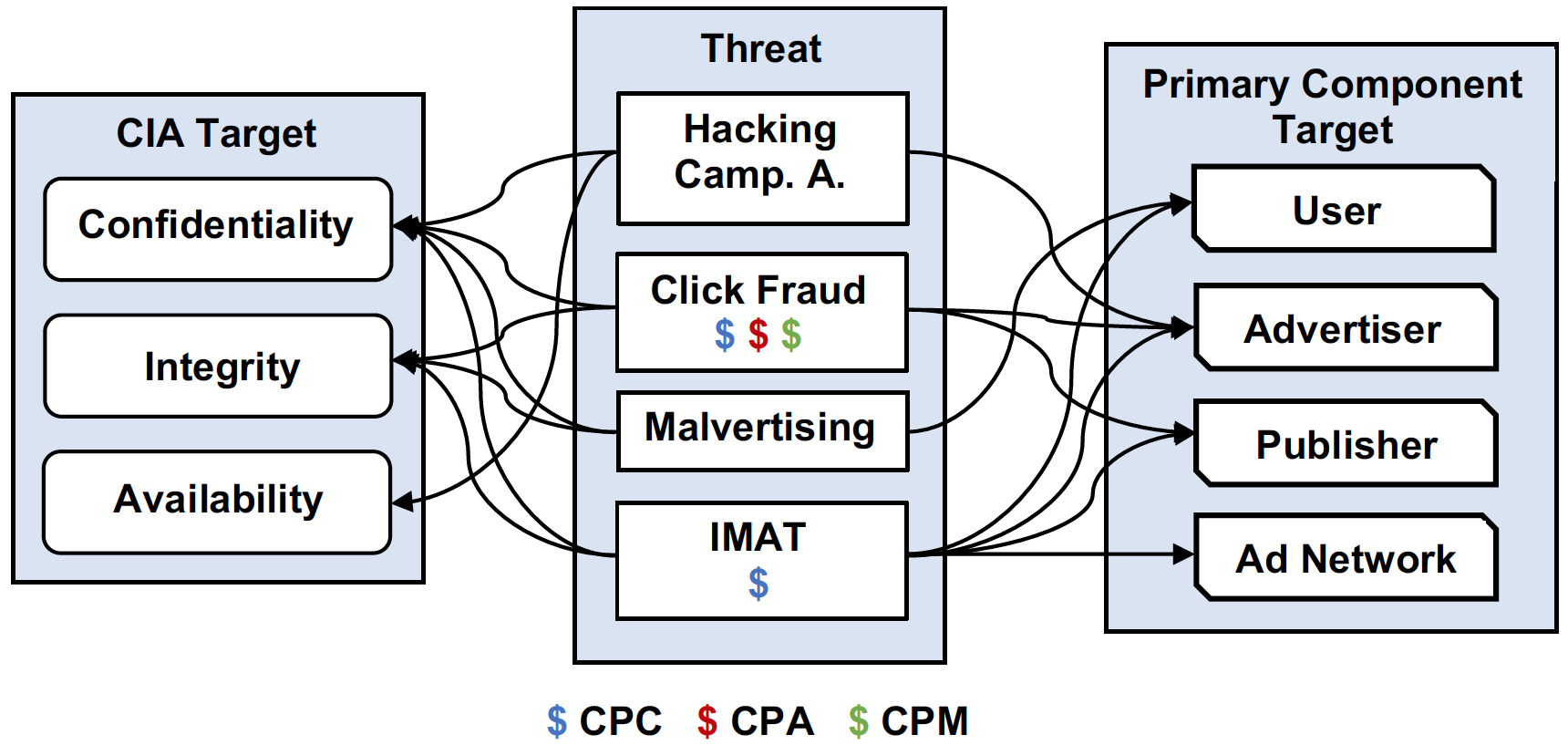}
		\caption{\small The linkage between threats, CIA target, and primary component target. CPC, CPA, and CPM are the revenue models. IMAT := Inflight Modification of Ad Traffic, Camp. A. := Campaign Account.}\label{fig:txonomy_relation}
	\end{figure*}
	
\subsection{Putting All Together}
\label{sec:compare}
This section explains, for each type of attack, how an adversary can exploit the risks in online advertising systems and conduct ad fraud. We review the methods that fraudsters use to gain money from online advertising systems, present a brief analysis of the goals of these attacks, and identify which revenue model is the adversary's goal and which components of the online business system could be the primary targets. To aid in comprehension, the results of this comparison are presented in Table~\ref{table:tbl_attack}.
	
The technical reasons for the threat are mainly due to aspects of human or professional weakness or negligence. In general, by applying regulatory frameworks in the industry, at least some security issues can be addressed. However, due to cost and lack of awareness, the online advertising security framework does not seem sufficient to combat threats.
	
In Fig.~\ref{fig:txonomy_relation} , we consider three dimensions to identify the relationship between threats, which players in the system are the attacker's target, which revenue models can be affected by the threats, and the CIA target. For example, click fraud can affect advertisers and publishers in the system. It can also compromise the confidentiality and integrity of the system. All revenue models, including CPC, CPA and CPM, can be affected by click fraud.
	
Among the attack, only Inflight modification of ad traffic attack can be applied to all components of the ecosystem. The hacking campaign account and malvertising only affect the advertiser and the user, respectively. Click fraud can endanger both the advertiser and publisher. Among the CIA principles shown on the left side of Fig.~\ref{fig:txonomy_relation}, only confidentiality can be compromised by all threats. Hacking can affect availability of the system. The integrity of the system can be compromised by click fraud, malvertising, and IMAT. All the revenue models can be infected by click fraud, while Inflight modification of ad traffic can only impact on CPC model.

\section{countermeasures for online advertising attacks}
\label{sec:count_mea}

%

In this section, we discuss several approaches proposed in the literature to combat various types of attacks on online advertising systems. Table~\ref{table:tbl_conuterm} summarizes the existing detection methods for online advertising systems and gives a preliminary overview of the pros and cons of using these methods. 

\subsection{Countermeasures to Hacking Campaign Account}
\label{sec:cntmea_hacking}

When Google AdWords~\cite{GOOGLECompany} was launched in 2000 and quickly became Google's primary source of revenue, it soon became a rich source of targets for ad fraud attacks.

Various reports and forums have discussed the fact that the majority of Gmail address and passwords are used to hack campaign accounts. The different approaches used to hack Google AdWords accounts can be categorized as $(i)$ brute force login attacks; $(ii)$ email spoofing; and $(iii)$ malware and spy tools for obtaining user account information~\cite{LEFTYGBALOGH}. When fraudsters enter to a campaign account, they can duplicate campaigns. Attackers can also generate enormous numbers of clicks and redirect destination URLs to other companies~\cite{MOZBlog, GOOGLE}.

One of the more straightforward options for preventing this type of attack is to select strong passwords. The security and protection of an account can be increased by choosing a complicated and lengthy password combining letters and numbers with special characters, and by changing the password regularly. It is also possible to monitor and control browsers with phishing filters~\cite{GOOGLEE}, especially when connecting to unsecured WIFI connections and signing in to Google accounts. Industries and business owners should have a contingency advertising plan for monitoring their revenue trends to handle the drop in revenue caused by fraudulent campaigns.

To detect hacking attacks in online advertising, a daily check can be carried out of accounts to guarantee not only the cost-benefit ratio and performance of the campaign but also to protect the campaign from reputational damage and loss of income. It is vital to monitor and analyze the performance of each AdWords campaign on a daily basis~\cite{mladenow2015online}.

\subsection{Countermeasures to  Click Fraud}
\label{sec:cntme_clickF}

In this section, we first discuss countermeasures for crowd fraud and then countermeasures for conventional ad fraud.

\subsubsection{Countermeasures to Crowd Fraud}
\label{sec:cntme_crowd}


The techniques typically used in business markets for crowd fraud detection mainly emphasize human interactions, including prior knowledge of malicious queries and principles associated with filtering. These approaches are costly, and tend to become invalid quickly because web workers may change their patterns of behavior to avoid detection.

To address these problems, the authors of~\cite{tian2015crowd} investigated the group behaviors associated with crowd fraud, and found that compared with the individual actions of each worker, which may involve considerable noise, group behaviors were more continuous.

In formal terms, these authors discovered certain typical feature distributions and network functions of crowd fraud that can be effectively applied to detect this activity. They noted the following aspects: $(i)$ moderateness: crowd fraud sometimes targets advertisers or queries with medium hit frequencies; $(ii)$ synchronicity: Internet users participating in crowd fraud can classify into coalitions~\cite{49zhang2016detecting} via which they typically target a distinct collection of advertisers and execute the fraud quickly; and $(iii)$ dispersivity: surfers involved in crowd fraud may search for an irrelevant series of topics and click advertisements from different industries simultaneously.

Based on the attributes mentioned above, the authors of~\cite{tian2015crowd} introduced an efficient solution for crowd fraud in search engine advertising, which was divided into three phases: \emph{constructing}, \emph{clustering}, and \emph{filtering}. In the constructing phase, they deleted irrelevant data from raw data logs of queries that did not meet the moderateness condition (e.g., either markedly small or large hit frequencies) to create a surfer-advertiser bigraph in which each edge referred to a single unique click history and included aspects such as search queries and hit times. Then, they built a surfer-advertiser inverted list for this bigraph for the next phase. In this list, each entry referred to the click history for each unique surfer. In the clustering phase, they described the sync-similarity between click histories to discover coalitions of surfers, indicating synchronicity.

Next, they converted the coalition detection system into a clustering problem that could be solved through a nonparametric clustering algorithm (such as DP-means~\cite{kulis2011revisiting}). After the clustering phase, the percentage of finding coalitions was high, and this caused false detections and therefore false alarms. For instance, in some business domains such as healthcare or games, regular Internet users with related interests may repeatedly click on the same advertisements to receive similar services. Hence, using infiltering, they created a filter for clusters based on the dispersivity to eliminate false alarm clusters.

Since this method does not require tuning of any parameters, it can be applied in real scenarios to find an infinite number of coalitions without human interaction. The authors also built a parallel version of their detection method (by parallelizing the nonparametric clustering algorithm) to make the system more scalable for massive web searching. The results of this experiment validated the accuracy and scalability of their approach. Although the proposed algorithm was capable of detecting crowd fraud, however, it failed to prevent this fraud. Moreover, evaluating the accuracy of the algorithm was hard due to the difficulty of collecting fraud data.

The crowd-sourcing click fraud detection model in~\cite{jiarui2015detecting} was based on a clustering analysis. In order to analyze the data, the study defined the distinct features of denseness, moderateness and concentricity. The model consisted of three phases: preprocessing, group detection and post-processing. In the preprocessing phase, queries that were less likely to be fraudulently clicked were removed. In the group detection phase, a crowd-sourcing click fraud group was treated as equivalent to a cluster, and a DPMeans clustering method was used to detect malicious groups. In the post-processing phase, demand clicks checked by mistake were filtered. Although the model achieved good convergence and extensibility, the complexity of this method was excessively high.

A novel crowdsource based system presented in~\cite{mouawi2019crowdsourcing} called Click Fraud Crowdsourcing (CFC) addresses the problem of crowd fraud in the mobile system by protecting both advertisers and ad networks. The method consists of four components: ad banners, ad network APIs, advertisers' website, and CFC component, which acts as a click fraud detection engine. The program can monitor the user's duration on the advertiser's website and simultaneously collect request data for several ads related to the different ad network, publisher and advertiser. The results showed the proposed method when detecting click fraud, compared to the solutions offered in the literature, while maintaining a high true positive rate (0.9), offers a false positive rate (0.1).

\subsubsection{Countermeasures to Conventional Ad Fraud}
\label{sec:cntme_conveF}

Countermeasures for different types of conventional Ad fraud discuss in the following.


\begin{itemize}[leftmargin=*]
	\item \textbf{Countermeasures to Badvertising.} A successful badvertisement stealthily and artificially generates automatic clicks on advertisements when users visit a site hosted by a fraudster, and can persist unseen by auditors from the ad provider. It does not require any specific technical knowledge to run this kind of attack, and any illegal webmaster can perform it~\cite{gandhi2006badvertisements}. 

At first glance, it may seem easy to detect this attack by controlling the CTR from the intended domain, but this is not always the case. For example, the attacker can generate both click-throughs and non-click-throughs by manipulating the traffic in the damaged page, while the customers correlated to those types are not informed of the advertisement. It should be noted that the owners of the site who earn income for a ``badvertiser''  may not be aware of their participation in running the attack. For example, the owners of a domain may be pretending not to know of the existence of an attack, or may be fooled by a corrupt webmaster. The former case corresponds to a phishing attack~\cite{tripathi2017novel}.

Developing tools for the discovery and prevention of frequent click fraud attacks is a major aim of industries in this field. AdWatcher~\cite{AdWatcher} and ClickProtector~\cite{ClickProtector} are two well-known companies that try to detect and prevent such attacks. The most common attack types are malware-based, which use automated scripts, individuals hired to deplete their competitors' advertising budget~\cite{India} or proxy servers to generate fake clicks. These attacks can be detected by tracking the IP addresses of the systems that generate the clicks or by distinguishing the click registered domains. Companies try to identify aspects such as duplicate clicks for a specific ad by a single IP address or irregularities in the traffic history, and to carry out careful analyses. However, a badvertising attack cannot easily be detected using these approaches, and there is a pressing need for other types of mechanisms to detect and prevent this attack.

The countermeasures discussed in~\cite{gandhi2006badvertisements} involve the construction of a ad code to detect an attack when preventing it is not possible. These methods can be divided into two types: \emph{active} and \emph{passive}. Active methods are used to detect click fraud, while passive methods are used to monitor the progress of a click fraud.

In formal terms, an active client-side solution is based on interactions with search engines, the execution of public searches, and visits to the resulting sites. It can carry out web surfing in a manner similar to the user. An active mechanism can conceal its status such that an agent cannot recognize it as a robot, and can present itself as a real user to the servers in order to interact with the agent and other entities.

In contrast, passive client-side approaches monitor the actions performed by users that lead to a click. It is possible to trap requests for advertisements by virtual execution of JavaScript code, and any attempt to display a specific web page in a way that it should be occurring after a click can be considered a fraudulent request. It should be pointed out that although this solution can be used against automatic click-fraud, it cannot be applied to protect a system against a type of attack that first creates a significant delay and then performs a click fraud. The only way to do this and to capture a delay is to let the virtual machine randomly select scripts for generating a delay.

We should recall that long delays are not preferred by attackers, since their session might be disconnected from the target before they can generate a click on the website. Passive client-side methods can be included with security toolboxes or anti-virus programs.

Another form of passive scheme is an infrastructure component. That can detect click fraud by shifting traffic, identifying candidate traffic and mimicking the system of the user receiving the packets. Example applications of infrastructure component schemes include an ISP-level spam filter and Mail Transfer Agent (MTA).
 
We can conclude from a performance analysis that if a client-side detection mechanism is installed only by a small proportion of customers, these attacks become entirely unprofitable.


\item \textbf{Countermeasures to Hit Shaving.} The author of~\cite{Johnson} explained that the rationale behind all inflation and deflation fraud (also called hit shaving) is a lack of knowledge. In both attacks, the entities who perform the fraud may under- or over-count transactions for financial gain, and it is difficult for the victim to prove the damage that arises. As a consequence, a general technique for detecting these frauds is to collect information relating to the victim's claim.

For example, in the case of deflation fraud, the authors of~\cite{Johnson} proposed the use of an online Trusted Third Party (TTP) as a mediator to facilitate interactions between two parties. To detect deflation fraud, the publisher must collect as much information as it can, based on the advertiser's claim. In a nutshell, the more info a publisher can gather, the stronger the detection scheme. The disadvantage of this solution is that it cannot be applied to the online advertising ecosystem. Similarly, in Google's AdWords, the publisher directly monitors the transactions. The methods mentioned above suffer from a lack of scalability and efficiency, since the publisher can interfere in the business operation of the advertiser and in turn with the TTP.

In~\cite{ding2010hybrid}, an efficient and flexible mechanism was proposed to relax the security solution slightly. The authors point out that there is a certain level of tolerable counting error for the publisher if they miss some transactions. Their mechanism involved a novel deflation fraud detection scheme that applied cryptography and probability-based techniques with the following features: $(i)$ the publisher can detect deflation fraud with a high probability of success, and the security parameters can be tuned by the publisher to provide a balance between cost-effectiveness and security assurance; $(ii)$ under these conditions, the web publisher can estimate and detect the expected number of transactions on a large scale; $(iii)$ although a transaction takes place only between advertiser and users, the proposed scheme is easy for end-users, since they are not required to keep any secret information; $(iv)$ the costs (such as computation, communication, and storage) of this method are all constant, making the scheme efficient and scalable.

The proposed hybrid method does not require the cooperation of a third party, and retains the simplicity of the current advertising system. The publisher also has the option to tune the security parameters to balance the security and cost of the model. The drawback of the proposed scheme is the need for manual tuning of parameters by the publisher.

Although there are many click-through payment mechanisms on the web, the publishers cannot verify whether they have received payment for each click-through to the target site. This allows for hit shaving, in which the target sites can avoid paying the publisher sites for some click-throughs.

The study in~\cite{reiter1998detecting} proposed some rapid and straightforward approaches to enable referrers to track the number of click-throughs, allowing them to be aware of how much money they are owed. These methods included ways of creating web pages and Common Gateway Interface (CGI) scripts that offer the referrer webmasters a greater ability to monitor the numbers of legal clicks, and also which pages the users click. They implemented these approaches by placing upper bounds and lower bounds on referrals. These are effective techniques that do not require awareness or cooperation by the webmasters of the sites to which the referrals are made.

The authors also explore more aggressive approaches for cooperating with the providers of click-through mechanisms, to allow webmasters to more accurately control the number of click-throughs. Although this second group of approaches requires cooperation by the webmasters of the click-through payment programs, it does not need trusted webmasters, since any failure to cooperate is quickly detectable. This is a robust solution: a referrer can discover this fraud after 20 times probe even if the target shaves only 5\% of the commission. However, this method is not always feasible, for example if the target website sells expensive items. In this method, referrers are expected to report their payments for leads and sales correctly, with the help of the target sites. Although techniques presented here are mainly invisible to the web user, their main disadvantage is the communication overhead for implementing the protocol, which causes it to be an inefficient and inflexible scheme.


\item \textbf{Countermeasures to Hit Inflation.} Due to the nature of hit inflation attacks, they are an important concern for advertising commissioners~\cite{zeller2004each}. Most research to date has focused on publisher fraud, since this can also be generalized to advertiser fraud. In the following, we therefore concentrate on publisher fraud unless it is specifically necessary to investigate advertiser fraud. We start with examples of classical approaches to inflation fraud detection, and give an overview of cryptography-based methods. Finally, we argue that the application of statistical analysis to streams of traffic is the most appropriate way to detect hit inflation.




\begin{enumerate}[leftmargin=*]
	
	\item \textbf{Classical Approach.} Classical fraud detection, also called offline fraud detection, employs a variety of metrics to evaluate publishers according to the quality of traffic to their websites~\cite{metwally2007hit}. The quality of traffic can be measured by its adaptation with normal network traffic. In classical detection methods, brokers can store the total traffic in databases and validate the quality (based on certain metrics) of the stored traffic using complex SQL scripts.

One of the most appropriate metrics is the CTR of the advertisements, which is constant across websites of the same type~\cite{klein1999defending}, while advertisements of different types have different CTRs on identical sites. If the website automatically visits and clicks, consequently, not only produce similar CTRs for the advertisements but rather the CTR of the displayed advertisements deviates from the normal values. Commissioners can develop this technique to monitor the behavior of advertisements by loading empty advertisements into the websites of publishers and checking clicks on these false advertisements.

However, classical metrics have several problems. They are not efficient metrics, since fraudsters can easily circumvent traditional tools, and can fool classical detection tools by abusing the site architecture~\cite{metwally2006hide} of a specific publisher to model the network metrics of advertisements and gain information about the parameters of the advertisements displayed on their website.

A lack of scalability is the second problem. It should be noted that the average impressions per second currently received by the commissioner is 20K, corresponding to 70M records that need to be stored in a database per hour. It is clear that executing SQL scripts to compute these metrics will lead to a decrease in database performance, and commissioners therefore execute them only periodically. Moreover, the updating of these metrics is also not scalable. Each click on an ad in any site may mean that the statistical parameters and the ranking of the website need to be recalculated.

Thirdly, the classical approach was developed before Internet advertising reached maturity, and hence represents the standard conflict between advertisers and commissioners. Traffic that does not adapt with the network metrics may be legal, although it will be low-quality traffic. Since classical methods are unable to detect malicious intent, they omit legitimate traffic with low quality.


	\item \textbf{Cryptographic Approach.} There are various cryptographic methods in the literature that can replace off-line measures. The central idea behind these is to change the industry standard to give fraudulent publishers less chance to conduct fraud~\cite{naor1998secure}. For example, in~\cite{jakobsson1999secure}, a simple model involving e-coupons was developed. In this model, the advertiser exploits cryptographic algorithms to produce coupons and distributes them to the publishers. Then, the publishers redistribute the coupons to users, who can use these cryptographic coupons to purchase items from the websites of advertisers. Web advertisers favor this model because it is based on pay-per-sale. Most publishers prefer to be paid based on the number of clicks or impressions, since this relates to the load on their servers.

Conversely, advertisers can exploit the model to receive a vast amount of clicks or impressions, which are essential to increase awareness of their brand. The authors claim that the proposed model meets most security and safety requirements; however, the model is vulnerable to hit shaving attack by advertisers.

The solution proposed by Goodman~\cite{goodman2005pay} is to replace the current pay-per-click scheme used in online advertising with a pay-per-impression system. This approach does not involve a monetary cost to the advertiser for click fraud, since they are no longer paid per click. The authors of~\cite{juels2007combating} suggest a cryptographic technique for changing the CPC model to CPA in which valid clicks are identified rather than invalid clicks being removed. This model guarantees the legitimacy of the clicks received by advertisers through a TTP. However, this model requires sharing information between third parties, which is not possible due to the security restrictions in modern browsers.

Other cryptographic methods rely on assistance from users to identify fraudulent traffic from regular traffic. Different groups of protocols using basic cryptography methods have been introduced to count the total number of visitors viewing a website~\cite{reiter1998detecting, blundo2002sawm}. One framework requires users to register with a broker, from which the user receives a token from the broker to use free services on the website of the publisher. The broker also shares the corresponding token with the publisher to allow them to recognize registered users. In this way, each time a user visits the publisher's website and sends a token-based authentication to the publisher, access is granted to that free service. The user updates the publisher via a hash function when an authentication token is sent to the publisher. Since publisher cannot predict the cost of the next visit (but can verify the value of the token), the number of user visits stored in the last token is sent back to the publisher at accounting time.

There are some limitations to this framework. Firstly, it presumes that the users trust brokers to download code to run the hash function~\cite{khare1998trust} and communicate with the publishers' servers. Secondly, it suffers from a lack of scalability, since numerous hash functions are required (one for each user). Thirdly, this scheme needs brokers to identify users uniquely in order to be effective, although exposing personal information on the users to the brokers violates the user's privacy. The last problem can be handled by user registration in the broker's website (by exposing the user's personal information). Brokers can also track and monitor the behavior of users by downloading spyware~\cite{saroiu2004measurement} onto their systems.


	\item \textbf{Data Analysis Approach.} Many advanced data analysis technologies have been developed to alleviate the problems caused by cryptographic methods. The principal aim of these technologies is to find particular patterns that characterize fraudulent traffic. As mentioned above, a broker needs to deal with the conflict between protecting the user's privacy and security, and the best way to address this challenge is to carry out statistical analysis on collected data (such as cookie IDs and IPs) with the help of temporary user identification. 
	

Cookies do not store any personal information, and the user has the ability to block, accept, or periodically clear them~\cite{mcgann2005study}. IP addresses can also be assigned to the user temporarily, and can be shared with other users. There is therefore no reason to change the industry model and to obfuscate the identity of the users when applying data analysis methods to cookie IDs and IPs, and these methods can detect fraud with high accuracy~\cite{metwally2007detectives}. The known data analysis approaches to defending against hit inflation are described below.


\textbf{Detecting duplicate clicks.} Since some publishers try to increase the number of clicks on their websites by clicking the same advertisement, some detection techniques rely on searching for duplicate clicks in the clickstream~\cite{zhang2008detecting, metwally2005duplicate}. The detection of duplicate clicks within a short time (for example single a day) raise suspicion for the commissioner.

In classical data analysis techniques, the commissioner can store the total traffic in databases and run complex SQL scripts to find duplicate clicks within a certain period. However, this method suffers from scalability and performance problems. Storing traffic in the database and then checking them to find duplicate clicks is very expensive for commissioners, since they receive a vast amount of traffic (an average size of around 70M records is generated per hour). In a online scenario, a detection scheme also needs to be fast, and should process the total traffic entry within 50$\mu$s. Hit inflation detection is therefore a critical part of streaming and sampling algorithms.

To cope with the above problem, Metwally et al.~\cite{metwally2005duplicate} proposed a fast algorithm for detecting duplicate clicks in data streams. Their algorithm relies on original Bloom filters~\cite{bloom1970space} and aims to find click fraud with an error rate of less than 1\%. They provide different solutions by considering three types of window, as follows: sliding windows (finding duplicate clicks corresponding to the last observed part of the stream); landmark windows (keeping particular parts of the stream for deduplication); and jumping windows (a trade-off between the first two types).

The results of an experiment on a real dataset show that within one day, one ad was clicked 10,781 times by users with the same cookie ID. Since the method is successful in identifying fraudulent intent, it can be considered a complementary approach to classical schemes that cannot differentiate low-quality from malicious traffic. However, the method has high computational complexity of order O(n), since it needs to keep active click identifications in its memory until they expire.

To address this problem, two algorithms, namely the Group Bloom Filter (GBF) and Timing Bloom Filter (TBF) algorithms, were developed in~\cite{zhang2008detecting}. The difference between them lies in the number of sub-windows. The GBF can detect click fraud using jumping windows with a small number of sub-windows, whilst TBF achieves this using a large number of sub-windows. These two algorithms involve simple operations and relatively little storage space, with zero false negatives. The error rate of duplicate detection is also reduced to less than 0.1\%.

Recently, with the development of ML, this approach has been leveraged to find patterns of difference between fraudulent click data streams. Some researchers have claimed that abnormal click-stream traffic is often a simple reuse of legal data traffic, and have tried to identify click fraud by detecting repeating patterns in a given click-stream for an ad. The Clicktok tool used a Non-negative Matrix Factorization (NMF) algorithm to partition click traffic to identify fraudulent clicks~\cite{nagaraja2019clicktok}. The authors claimed that the proposed solution reached an accuracy of 99.6\%. Despite this high efficiency, however, the solution only works on the client-side. A deep learning-based model was used in~\cite{thejas2019deep} to build a multi-layered neural network with an attached autoencoder and generative adversarial network to detect click fraud. The authors of~\cite{jiang2021generating} proposed a new way to identify inherently hidden patterns performed by fake or malicious users on ad networks. They argued that training machine learning models directly on fraudulent and benign sequences collected from advertising activities were not easy. The majority of ad traffic is non-fraudulent, and data labelling by a human is time-consuming. Hence, they combined a variant of Time-LSTM cells combined with a modified version of Sequence Generative Adversarial Generative (SeqGAN) to generate artificial sequences to impersonate the fake user patterns in ad traffic. They also reduce computational costs by using Maximum Likelihood Estimation (MLE) pre-training and a Critic network for seqGAN training. They claimed that the use of sequences generated by GAN could increase the ability to classify event-based fraud detection classifiers.

Despite the effectiveness of ML methods for detecting fraudulent clicks, because large amounts of data and sophisticated fraudulent ads are constantly involved in the online advertising system, these methods may suffer from labour-intensive feature engineering and the power of detection algorithms. Also, an attacker could easily modify their fraud patterns based on existing fraud detection characteristics and rules to prevent identification. Hence, many studies have focused on graph-based methods for detecting anomalies (e.g., fraud detection) in systems~\cite{pourhabibi2020fraud}. The study in~\cite{hu2020gfd} proposed a weighted heterogeneous graph embedding and deep learning-based fraud detection method (called GFD) to detect fake applications for mobile advertising. The proposed method has three steps: $(i)$ use a weighted heterogeneous graph to display behavioural patterns between users, mobile apps, and mobile ads, and design a weighted meta path to vector algorithm to learn node representations (graph-based features) from the graph; $(ii)$ use a time window-based statistical analysis method to extract attribute-based features from the sample table data; and $(iii)$ propose a hybrid neural network for fuse graph-based features and attribute-based features classifying fake apps from normal apps.

In~\cite{zhang2018click}, a cost-sensitive Back Propagation Neural Network (CSBPNN) architecture was implemented for click fraud detection, and the researchers applied an Artificial Bee Colony (ABC) approach to optimize the CSBPNN connection weights and feature selection. However, many types of click fraud do not rely on legitimate streams of click data, and attackers are also able to construct data streams with patterns similar to legitimate click data streams through the use of fitting classifiers. In this case, the performance of the above systems will be significantly degraded.
\end{enumerate}
\end{itemize}

\textbf{Fabricated impressions and clicks.} Other solutions collect ad traffic across user IPs and cookie IDs to identify fabricated clicks and impressions. They are based on finding client behavior (e.g., advertisement traffic) that deviates from normal behavior~\cite{metwally2007detectives, metwally2007hit}.

Cryptographic and classical methods cannot determine the difference between attacks launched by a single publisher and by a group of publishers (also called a coalition attack). In principle, making this difference is the main idea behind the data analysis approach. Although it is easy for coalition attacks to defraud classical methods, data analysis mechanisms have been developed to try to find evidence of these attacks~\cite{metwally2007detectives}. 


Metwally et al.~\cite{metwally2005using} designed a scheme to detect the hit inflation attack identified in~\cite{anupam1999security}. They observed that several websites could cooperate to make fake clicks and consequently improve their business interests, and proposed an algorithm named \emph{Streaming-Rules} to detect hit inflation in an online advertising system. This approach relies on discovering the association rules (defined as forward and backward association rules) between each pair of corresponding elements in the stream. This algorithm requires cooperation between ISPs and brokers. An ISP can recognize which websites are generally visited before a particular website, while maintaining users' privacy~\cite{iqbal2018protecting}, by analyzing the entire HTTP requests stream. The authors claimed that Streaming-Rules could discover the association between elements occurring in a stream with tight error guarantees and minimal memory usage.

The solution proposed in~\cite{metwally2005using} is not efficient against other coalition attacks, since it is designed to detect the specific attack described in~\cite{anupam1999security}. For example, if each adversary in the coalition attack takes control of the user's system via Trojans, then the adversary can separate the HTTP request stream by ISP, making it impossible to detect the attack using Streaming-Rules. Hence, in~\cite{metwally2007detectives}, an approach was developed to identify different types of sophisticated coalition attacks (e.g., a coalition formed of multiple dishonest publishers) called the Similarity-Seeker algorithm. This detection mechanism relies on analyzing traffic to find similarities in the traffic to websites. Legitimate websites do not have similar traffic, and traffic from similar sets of IPs is therefore suspicious. The original model can discover coalition attacks of size two, and the extended model can find attacks by coalitions of arbitrary sizes. The exploitation of statistical traffic analysis gives more scalability than traditional technologies.

Another method presented by Metwally et al. in~\cite{metwally2008sleuth} called Single-pubLisher attack dEtection Using correlaTion Hunting (SLEUTH) addresses the problem of fraudulent traffic generated by a single publisher via several IPs. This approach focuses on discovering an association between the publisher and the IP address of a machine. However, SLEUTH is only an adequate solution for a botnet that utilizes a vast number of IPs, and assumes that the traffic features of non-fraudulent publishers and IPs are constant. This assumption is not applicable to online advertising systems, where trends are highly temporal.

AdSherlock~\cite{cao2020adsherlock} is another method in this category in client-side that is used for online click fraud detection, and which is based on an ad request tree model. In this type of method, users need to install additional programs on their devices, which is not highly applicable. ClickGuard~\cite{shi2020clickguard} is an machine learning-based (ML) system for detecting click fraud attacks. This system uses a classifier that is trained based on the motion sensor signals in mobile devices, in order to separate benign clicks from fake ones. However, despite the high accuracy of this novel inference system, it is a client-side application.

Although these solutions have certain benefits, all of them are under the threaten of complicated botnet ad fraud~\cite{stone2011understanding}. Many compromised machines are used to modify the IPs and cookie IDs of fraudulent requests.

In~\cite{haddadi2010fighting}, the authors described the use of bluff advertisements, an online click-fraud detection strategy that blacklists malicious publishers based on a predefined threshold. This approach was designed to display several unrelated/fake advertisements amongst the user's targeted advertisements, with the expectation that these advertisements will not be clicked on. In addition to monitoring IPs and applying profile-matching and threshold detection techniques, bluff advertisements can create some obstacles for botnet owners who want to train their software. Negative attitudes of users can also be reduced by decreasing the number of precisely targeted advertisements. These considerations motivated the authors of~\cite{dave2012measuring} to recommend a technique for advertisers to count the proportion of invalid clicks on their advertisements by generating fake ones. Running bluff advertisements leads to an increase in advertising budgets for advertisers.

A hybrid ML method called Cascaded Forest and XGBoost (CFXGB) proposed in~\cite{thejas2021hybrid} to identify faulty clicks. The proposed model combines two learning models for feature transformation by Cascaded Forests and classification by XGBoost. They have shown that this combination leads to better results compared to using only cascaded forests as classifiers. Despite the high accuracy of the model, it is necessary to adjust the parameters.

All of the above detection methods can only address fraud after it has occurred. The authors of~\cite{costa2012proposal} therefore proposed a new automated method for preventing click fraud called clickable Completely Automated Public Turing test to tell Computers and Humans Apart (CAPTCHAs). In the proposed method, customers complete a simple Turing test~\cite{turing2009computing} and are then diverted to the publisher's site. Although click fraud can be identified based on valid users, the loading of CAPTCHAs requires time and space. 

The techniques listed above are related to finding click fraud in web content. There are some other approaches to examine the detection and control of ad fraud in mobile applications. Based on the similarities that we mentioned in Section~\ref{sec:mobile_adv} regarding the web and mobile ad infrastructure, they face similar threats as click fraud~\cite{shaari2020extensive}. However, despite the similarities, the threats in these systems are in different forms. Ad threats behave differently on the two platforms because threatening features such as click fraud and malvertising in mobile platforms are different from web ads. For example, the mobile attacker can automatically generate fake ad clicks, while web attackers have to launch click fraud with bots. Based on these observations, the following are some of the methods aimed at detecting mobile click fraud.

The authors of \cite{aberathne2020novel} proposed a hidden Markov-based automated classification algorithm to detect fraud impressions in mobile advertising. The algorithm is called the hidden Markov scoring model (HMSM), which is based on the hidden Markov model (HMM) scoring approach instead of a conventional HMM probabilistic model. Although having a large data set in ML-based techniques helps achieve better accuracy, they claimed that the optimal size for training is 5000. Optimal selection of training sample size leads to reduced computation time in training and storage space.

MAdFraud~\cite{crussell2014madfraud} is dynamic testing frameworks that have focused on detecting click fraud on Android by analyzing network traffic. MAdFraud is designed to detect fake URL requests without valid user clicks. They first identify the unique features of mobile that leads to fraud. On the Android platform, at any time, an app runs in the foreground, where the app has a UI. Their first observation was that when an application receives ads in the background, it is probably fraud because the developer of the application receives the ad's credit without showing it to the user. The second one was that when an application clicks on an ad without user interaction, it is fake. They create a new way to detect ad impressions and clicks in three phases automatically: building HTTP request trees, identifying ad request pages using ML, and applying heuristics methods to detect HTTP request trees' clicks. Despite the efficiency of the tool in finding fraud apps, its observation is without user intervention, and it takes time to collect the app process.

MAdLife~\cite{chen2019revisiting} is another analytics tool in this category that detects the full-screen ad impressions used without any user interaction. This study shows a correlation between click fraud and malvertising and shows 18.36\% of malicious ads loaded with click fraud. This tool is developed to monitor the ad traffic generated throughout its lifetime in Android WebView. It first starts by logging the pre-click data in a database. After automatically performing ad clicks with lightweight UI automation, it stores the post-click traffic data in another table in the database. Finally, the tool compares tables to find equality to classify apps as malicious ones. This detection method only considers clicking on WebView and does not consider fraudulent activities that do not involve WebView.

The authors of~\cite{cho2015empirical} have created a mobile advertising click bot, called ClickDroid, that periodically clicks on mobile ads. It tries not to be detected by fraudulent mobile clicks on ad networks. To do this, it modifies the device's identifier every time it clicks on a mobile ad. ClickDroid evaluated 100-click fraud on eight popular advertising networks, and only two mobile ad networks detected traffic anomalies, indicating the ad network's inability to detect click fraud. The authors also proposed a method to detect click fraud for Android apps based on tracking the user's clicks from the touch sensor at the kernel level. To do this, they install a middleware framework to gather the sensor output and save it in a separate file. This report can then be used to verify the events generated by ClickDroid based on the difference between the human and software-generated clicks. Assuming that advanced bots cannot bypass the middleware, the system is still susceptible to click-farms.

  \begin{table*}[!htbp]
  	\caption{\small Comparison on existing detection methods in online advertising system. Camp. := Campaign, Ref. := Reference, Badver. := Badvertising.}\label{table:tbl_conuterm}
  	\centering
    \footnotesize{
\setlength{\tabcolsep}{6pt} 
\renewcommand{\arraystretch}{1} 
  		\begin{tabular}{|p{0.2cm}|p{0.2cm}|p{4cm}|p{4.5cm}|p{4cm}|p{0.2cm}|}
  			\hline
  			\multicolumn{1}{|c|}{\textbf{Attack}} & \multicolumn{2}{c|}{\textbf{Countermeasure}} & \multicolumn{1}{c|}{\textbf{Advantage}} & \multicolumn{1}{c|}{\textbf{Disadvantage}} & \multicolumn{1}{c|}{\textbf{Ref.}}\\\hline
  			
  			\multirow{3}{*}{\begin{sideways}{\textbf{\parbox{1.1cm}{Hacking Camp. Account}}}\end{sideways}}&\multicolumn{2}{p{5.5cm}|}{Daily checking of the user's account}&
  			\begin{itemize}[leftmargin=*]
  			\item[\cmark] Protecting from possible financial and reputation losses
  			\end{itemize}
  			&
  			\begin{itemize}[leftmargin=*]
  				\item [\xmark]Highly time-consuming
  			\end{itemize}
  			&\multicolumn{1}{p{0.2cm}|}{\cite{mladenow2015online}}\\\hline

  			& 	\multicolumn{2}{p{5.5cm}|}{Detection strategies based on human interactions}  & 
  			\begin{itemize}[leftmargin=*]
  				\item [\cmark ]Strong detection scheme
  			\end{itemize} & 
  			\begin{itemize}[leftmargin=*]
  				\item [\xmark]Labor costs
  				\item [\xmark]Becoming invalid quickly due to rapid change in web workers' behavior 
  			\end{itemize}
  			 & \multicolumn{1}{p{0.2cm}|}{\cite{tian2015crowd}} \\\cline{2-6}

  			\multirow{-4}{*}{\begin{sideways}{\parbox{1.9cm}{\textbf{Crowd Fraud}}}\end{sideways}}& \multicolumn{2}{p{5.5cm}|}{Substantial randomness solution based on the group behaviors}  & 
  			
  			\begin{itemize}[leftmargin=*]
  				\item [\cmark]Robustness, scalable, and reliable
  				\item [\cmark]No need to tune parameters manually
  				\item [\cmark]Applicable in real-world 
  			\end{itemize} & 
  				\begin{itemize}[leftmargin=*]
  				\item [\xmark]Fails in preventing fraud
  				\item [\xmark]Difficulty in evaluating the accuracy of the algorithm
  			\end{itemize}& \multicolumn{1}{p{0.2cm}|}{\cite{tian2015crowd}}\\\cline{2-6}
  			
  					& 	\multicolumn{2}{p{5.5cm}|}{Detection method based on clustering analysis}  & 
  					\begin{itemize}[leftmargin=*]
  						\item [\cmark ]Good convergence and extensibility
  					\end{itemize} & 
  					\begin{itemize}[leftmargin=*]
  						\item [\xmark]High complexity
  					\end{itemize}
  					& \multicolumn{1}{p{0.2cm}|}{\cite{jiarui2015detecting}} \\\cline{2-6}
  					
  						& 	\multicolumn{2}{p{5.5cm}|}{CFC}  & 
  						\begin{itemize}[leftmargin=*]
  							\item [\cmark ]Offer crowdsourcing fraud tool to protect both ad networks and advertisers in mobile platform
  						\end{itemize} & 
  						\begin{itemize}[leftmargin=*]
  							\item [\xmark]Requires the advertisers and ad networks to trust the CFC party
  						\end{itemize}
  						& \multicolumn{1}{p{0.2cm}|}{\cite{mouawi2019crowdsourcing}} \\\hline
  			
  			\multirow{-1}{*}{\begin{sideways}{\textbf{\parbox{1.1cm}{Badver.}}}\end{sideways}}&\multicolumn{2}{p{5.5cm}|}{Detecting and preventing badvertisment via active and passive schemes}&
  			\begin{itemize}[leftmargin=*]
  				\item [\cmark]Preserving user privacy
  			\end{itemize}&
  			\begin{itemize}[leftmargin=*]
  				\item [\xmark]Needs third-party interaction 
  				\item [\xmark]Time-consuming
  			\end{itemize}&\multicolumn{1}{p{0.2cm}|}{\cite{gandhi2006badvertisements}}\\\hline

  			& \multicolumn{2}{p{5.5cm}|}{Collecting information} & 
  			\begin{itemize}[leftmargin=*]
  			\item  [\cmark]Strong detection scheme 
  	     	\end{itemize} & 
  				\begin{itemize}[leftmargin=*]
  			     	\item [\xmark]Lack of scalability 
  			     	\item [\xmark]Lack of efficiency 
  				\end{itemize}& \multicolumn{1}{p{0.2cm}|}{\cite{Johnson}} \\\cline{2-6}

  			& \multicolumn{2}{p{5.5cm}|}{
  				Using cryptography and probability tools to detect fraud}  & 
  			\begin{itemize}[leftmargin=*]
  				\item [\cmark] User-friendly and simple model
  				\item [\cmark] No need third party
  				\item [\cmark] Constant ad's communications, computation, and storage cost 
  			\end{itemize}
  			& 
  			\begin{itemize}[leftmargin=*]
  				\item [\xmark]Need to tune parameters manually
  			\end{itemize}
  				& \multicolumn{1}{p{0.2cm}|}{\cite{ding2010hybrid}} \\\cline{2-6}

  			\multirow{-3}{*}{\begin{sideways}{\textbf{\parbox{1.6cm}{Hit Shaving}}}\end{sideways}}& 	\multicolumn{2}{p{5.5cm}|}{Enabling the referrer webmasters to monitor the number of legal clicks}  & 		
  			\begin{itemize}[leftmargin=*]
  				\item [\cmark]No need awareness or cooperation by the webmasters
  					\end{itemize}
  				&  \begin{itemize}[leftmargin=*]
  					\item [\xmark]Communication overhead
  				\end{itemize}
  				 & \multicolumn{1}{p{0.2cm}|}{\cite{reiter1998detecting}} \\\cline{2-6}
  			
  			& \multicolumn{2}{p{5.5cm}|}{Enabling the providers of click-through mechanisms to control the number of clicks}  &	
  			\begin{itemize}[leftmargin=*]
  				\item [\cmark]Robust
  				\item [\cmark]No need to honest webmaster
  			\end{itemize} & 
  			\begin{itemize}[leftmargin=*]
  				\item [\xmark]Cooperation or awareness by the webmaster
  			\end{itemize}&\cite{reiter1998detecting} \\\hline

  			&\multirow{3}{*}{\begin{sideways}{\textbf{\parbox{1.41cm}{Classical}}}\end{sideways}}&Using a variety of metrics to monitor the quality of the traffic to find fraud&
  			
  			\begin{itemize}[leftmargin=*]
  				\item [\cmark]No need third party \end{itemize}&\begin{itemize}[leftmargin=*]
  				\item [\xmark]Lack of efficiency and scalability
  				\item [\xmark]Conflict of interest between commissioners and advertisers \end{itemize}&\multicolumn{1}{p{0.2cm}|}{\cite{metwally2007hit}, \cite{klein1999defending}, \cite{metwally2006hide}}\\\cline{2-6}
  			
  			&& Changing the industry model based on pay-per-sale& \begin{itemize} [leftmargin=*]\item [\cmark]Safe 
  				\item [\cmark]Robust \end{itemize}
  			& 
  			\begin{itemize}[leftmargin=*]
  				\item [\xmark]Vulnerable to hit shaving \end{itemize}&\multicolumn{1}{p{0.2cm}|}{\cite{jakobsson1999secure }}\\\cline{3-6}

  			&\multirow{3}{*}{\begin{sideways}{\textbf{Cryptographic}}\end{sideways}}& Changing the pay-per-click model with the pay-per-impression/ pay-per-action model& \begin{itemize}[leftmargin=*]
  				\item [\cmark]Guarantees the legitimacy of the receiving clicks by advertisers through a trusted third party\end{itemize}& \begin{itemize}[leftmargin=*]
  				\item [\xmark]Sharing the information between the third parties \end{itemize}&\multicolumn{1}{p{0.2cm}|}{\cite{goodman2005pay}, \cite{juels2007combating}}\\\cline{3-6}
  			
  			\multirow{7}{*}{\begin{sideways}{\textbf{Hit Inflation}}\end{sideways}}&& The assistance of the users to identify fraudulent traffic from regular traffic& \begin{itemize}[leftmargin=*]
  				\item [\cmark]Cost saving by free service \end{itemize}&
  			\begin{itemize}[leftmargin=*]
  				\item [\xmark]Lack of scalability and user privacy
  				\item [\xmark]Sharing the information between the third parties \end{itemize}&\multicolumn{1}{p{0.2cm}|}{\cite{reiter1998detecting}, \cite{blundo2002sawm}}\\\cline{2-6}

  			&\multirow{3}{*}{\begin{sideways}{\textbf{Data Analysis}}\end{sideways}}& Detecting duplicate clicks: 
  			\begin{itemize}[leftmargin=*]
  				\item Original Bloom Filter algorithm
  				\item GBF algorithm and TBF algorithm
  				\item Clicktok
  				\item Deep Learning-based Model
  				\item SeqGAN
  				\item GFD
  				\item CSBPNN-ABC
  			\end{itemize}
  			& \begin{itemize}[leftmargin=*]
  				\item [\cmark]Less error rate
  				\item [\cmark]Requires simpler operations and less storage space/ Low false-positive rate
  				\item [\cmark]Low latency
  				\item [\cmark]High accuracy
  					\item [\cmark]Quick classifier
  				\item [\cmark]No labour-intensive feature engineering
  				\item [\cmark]High accuracy
  			\end{itemize}& \begin{itemize}[leftmargin=*]
  			\item [\xmark]Memory waste
  			\item [\xmark] Theoretical analysis was made
  			\item [\xmark] Work on the user side
  			\item [\xmark] Low performance
  			\item [\xmark] -
  			\item [\xmark] Using one dataset for seven days leads to less robustness and accuracy
  			\item [\xmark] Low performance
  		\end{itemize}&\multicolumn{1}{p{0.2cm}|}{\cite{bloom1970space}, \cite{zhang2008detecting},  \cite{nagaraja2019clicktok}, \cite{thejas2019deep}, \cite{jiang2021generating},\cite{hu2020gfd}, \cite{zhang2018click}}

  		\\\hline

  		\end{tabular}}
  	\end{table*}

\begin{table*}
	\caption{\small Continued from Table~\ref{table:tbl_conuterm}.}\label{table:tbl_conuterm2}
	\centering
	\footnotesize{
		\setlength{\tabcolsep}{6pt} 
		\renewcommand{\arraystretch}{1} 
		\begin{tabular}{|p{0.2cm}|p{0.2cm}|p{4cm}|p{4.5cm}|p{4cm}|p{0.2cm}|}
			\hline
			\multicolumn{1}{|c|}{\textbf{Attack}} & \multicolumn{2}{c|}{\textbf{Countermeasure}} & \multicolumn{1}{c|}{\textbf{Advantage}} & \multicolumn{1}{c|}{\textbf{Disadvantage}} & \multicolumn{1}{c|}{\textbf{Ref.}}\\\hline
			
				\multirow{7}{*}{\begin{sideways}{\textbf{Hit Inflation}}\end{sideways}}&\multirow{3}{*}{\begin{sideways}{\textbf{Data Analysis}}\end{sideways}}& 	Fabricated impressions and clicks: \begin{itemize}[leftmargin=*]
					\item Streaming-Rules algorithm
					\item Similarity-seeker algorithm
					\item SLEUTH 
					\item AdSherlock
					\item ClickGuard
					\item Bluff Ads
					\item CFXGB
					\item CAPTCHAs
					\item HMSM
					\item MAdFraud
					\item MAdLife
					\item ClickDroid
				\end{itemize}
				& \begin{itemize}[leftmargin=*]
					\item [\cmark]Scalability and ability to detect specific hit inflation
					\item [\cmark]Highly scalable
					\item [\cmark]High accuracy \& ability to detect complex coalition attacks
					\item [\cmark]High accuracy \& Lower overhead
					\item [\cmark]High accuracy
					\item [\cmark]Put some obstacles against the botnet's owner to train their software
					\item [\cmark]Reducing computation time \& storage space
					\item [\cmark]High accuracy
					\item [\cmark]Identifying click fraud based on the valid user
					
					\item [\cmark]High efficiency
					\item [\cmark]Dynamic data analytics approach
					\item [\cmark]No specialized hardware requiredy
				\end{itemize}
				& 
				\begin{itemize}[leftmargin=*]
					\item [\xmark]Thwarted by sophisticated botnet ad fraud
					\item [\xmark]Under the threaten of complicated botnet ad fraud
					\item [\xmark]Not applicable to online advertising systems
					\item [\xmark]Work on the user side
					\item [\xmark]Work on the user side
					\item [\xmark]Increasing advertisers' budget on advertisements
					\item [\xmark]Less accuracy
					\item [\xmark]Needs to tune parameters
					\item [\xmark]Loading CAPTCHAs needs time and space
					\item [\xmark]Without user intervention \& Time-consuming
					\item [\xmark]Lack of trustability \& Applicable only for Android WebView
					\item [\xmark]Susceptible to click-farms attack
				\end{itemize} &\multicolumn{1}{p{0.2cm}|}{\cite{metwally2005using}, \cite{metwally2007detectives}, \cite{metwally2008sleuth}, \cite{cao2020adsherlock}, \cite{shi2020clickguard}, \cite{haddadi2010fighting}, \cite{thejas2021hybrid}, \cite{costa2012proposal}, \cite{aberathne2020novel},  \cite{crussell2014madfraud}, \cite{chen2019revisiting},  \cite{cho2015empirical}}
				\\\cline{1-6}
				
					& 	\multicolumn{2}{p{5.5cm}|}{Data integrity and authentication mechanisms}  & \begin{itemize}[leftmargin=*]
						\item [\cmark]Ensure the end-to-end security of communications to prevent inflight modifications \end{itemize}
					& 
					\begin{itemize}[leftmargin=*]
						\item [\xmark]Lack of scalability
						\item [\xmark]Highly communication cost \end{itemize} & \multicolumn{1}{p{0.2cm}|}{\cite{Rescorla}} \\\cline{2-6}

					\multirow{-4}{*}{\begin{sideways}{\textbf{\parbox{4.1cm}{Inflight Modification of Ad Traffic}}}\end{sideways}}& \multicolumn{2}{p{5.5cm}|}{Using a new encryption method to encrypt Web communications without other host authentication}  & 
					\begin{itemize}[leftmargin=*]
						\item [\cmark]Highly scalable \end{itemize} & \begin{itemize}[leftmargin=*]
						\item [\xmark]Fail to protect the system against MITM attacks \end{itemize}& \multicolumn{1}{p{0.2cm}|}{\cite{Langley}} \\\cline{2-6}

					& \multicolumn{2}{p{5.5cm}|}{Using Web Tripwire
						to detect inflight changes to websites}  & \begin{itemize}[leftmargin=*] \item [\cmark] A cheaper tool than HTTPS \end{itemize}& \begin{itemize} [leftmargin=*]\item [\xmark]Non-cryptographically secure method \end{itemize}& \multicolumn{1}{p{0.2cm}|}{\cite{Reis}} \\\cline{2-6}
					
					& \multicolumn{2}{p{5.5cm}|}{Secure scheme based on the collaboration between ad networks and web servers}  & \begin{itemize}[leftmargin=*]
						\item  [\cmark]Ensure authenticity and integrity of the traffic \end{itemize}& \begin{itemize}[leftmargin=*]
						\item [\xmark]Additional charge for publishers and ad networks \end{itemize}& \multicolumn{1}{p{0.2cm}|}{\cite{Vratonjic}} \\\hline

				\multirow{5}{*}{\begin{sideways}{\textbf{\parbox{2.2cm}{Malvertising}}}\end{sideways}}& 	\multicolumn{2}{p{5.5cm}|}{Checking the advertisements regularly and validate their appropriateness by publishers or ad networks} & 
				\begin{itemize}[leftmargin=*]
					\item [\cmark]Prevent losses of reputation, traffic, and revenue 
				\end{itemize}
				& 
				\begin{itemize}[leftmargin=*]
					\item [\xmark]Highly time-consuming \end{itemize}& \multicolumn{1}{p{0.2cm}|}{\cite{vratonjic2011online}} \\\cline{2-6}

				& \multicolumn{2}{p{5.5cm}|}{Install/update anti-malware software by users}  & 
				\begin{itemize}[leftmargin=*]
					\item [\cmark]Preventing to install malware on the user's machine \end{itemize}&
				\begin{itemize}[leftmargin=*]
					\item [\xmark]Use up a lot of memory \& disk space and slowing down the system \end{itemize}& \multicolumn{1}{p{0.2cm}|}{\cite{vratonjic2011online}} \\\cline{2-6}
				
						& \multicolumn{2}{p{5.5cm}|}{Applying the game-theoretic approach to formulating the malvertising problem}  & 
						\begin{itemize}[leftmargin=*]
							\item [\cmark]Data analytics approach \end{itemize}&
						\begin{itemize}[leftmargin=*]
							\item [\xmark]Theoretical analysis was made 
							\item [\xmark]Not applicable in real-world 
							\item [\xmark]Advertisers' benefits only 
							\end{itemize}& \multicolumn{1}{p{0.2cm}|}{\cite{huang2018bayesian}}\\\cline{2-6}
							
							& \multicolumn{2}{p{5.5cm}|}{Detection malvertising by ML}  & 
							\begin{itemize}[leftmargin=*]
								\item [\cmark]Detecting malvertising in web-based system  \end{itemize}&
							\begin{itemize}[leftmargin=*]
								\item [\xmark]Unable to detect unknown attacks
							\end{itemize}& \multicolumn{1}{p{0.2cm}|}{\cite{poornachandran2017demalvertising}} \\\cline{2-6}
							
							&
							\multicolumn{2}{p{5.5cm}|}{UI-based methodology to detect malware }  & 
							\begin{itemize}[leftmargin=*]
								\item [\cmark]Dynamic data analytics approach \end{itemize}&
							\begin{itemize}[leftmargin=*]
								\item [\xmark]Not applicable to find malware generating by ad network
							
							\end{itemize}& \multicolumn{1}{p{0.2cm}|}{\cite{suresh2019analysis}} \\\cline{2-6}
							
							& 
							 \multicolumn{2}{p{5.5cm}|}{MadDroid}  & 
							\begin{itemize}[leftmargin=*]
								\item [\cmark]Dynamic data analytics approach \end{itemize}&
							\begin{itemize}[leftmargin=*]
								\item [\xmark]Lack of trustability 
								\item [\xmark] Applicable only for mobile
							\end{itemize}& \multicolumn{1}{p{0.2cm}|}{\cite{liu2020maddroid}} \\\hline
						
%
%
		\end{tabular}}
	\end{table*}

%
%

\subsection{Countermeasures to Inflight Modification of Ad Traffic}
\label{sec:cntmea_InflighModi}

The authors of~\cite{vratonjic2011online} proposed data integrity and authentication tools to ensure end-to-end security for communication to prevent inflight modification. However, the use of these mechanisms has certain disadvantages that make them challenging to deploy on a wide scale. Firstly, authentication tools, such as Transport Layer Security (TLS) protocol, depend on cryptographic processes that impose a high computational cost on servers. In other words, the high security of TLS due to the complexity of the cryptographic algorithms adopted by TLS leads to high computational and energy costs~\cite{nofal2019comprehensive}. Secondly, since the authentication mechanism uses digital certificates to activate Web servers authentication, which are expensive since certificate authorities are required carry out authentication of web servers manually. Clearly, if a site has a certificate assigned by a trusted certification authority, a trusted connection can be made that helps browsers to authenticate websites~\cite{vratonjic2011online}. 

Web administrators also prefer to use a customized self-signed certificate without relying on third-party certification authorities to avoid the extra cost; however, such self-signed certificates are vulnerable to MITM attacks, and do not provide a reliable solution that allows the web browser to identify the website, and users need to decide whether or not to trust the corresponding website~\cite{Wendlandt}. From the user's point of view, it is complicated to determine the operation of a given certificate and to validate it. As a result, a malicious server can often communicate with users. A notary office can be established to control the consistency of the web server's public keys and to help the user verify self-signed certificates. Although this technique is a new and reliable solution, it has the same limitations as the scheme in~\cite{Rescorla}.

To tackle the above problems, researchers have introduced several alternative approaches to protect Web content effectively~\cite{Langley, Reis, Vratonjic}. For example, in~\cite{Langley}, the authors present a new opportunistic encryption method for encrypting web communications, involving a secure channel without other host authentication. However, this technique is unable to protect systems against MITM attacks, since the attacker can easily access the certificates used for authentication and replace them to impersonate web page. In other work, the authors of~\cite{Reis} adopted a web-based measurement tool called Web Tripwire to detect inflight changes to websites. This method can inject JavaScript code into the site and monitor the HTTP web page to identify any changes in it. The tool immediately reports any modifications to the web page to both the end-user and the web server. Tripwire is a cheaper tool than HTTPS, which checks the integrity of pages, but is a non-cryptographically secure method. In~\cite{Vratonjic}, a secure scheme based on a collaboration between ad networks and web servers was introduced to counteract inflight traffic modification. This method is based on the fact that ad networks with digital authentication certificates can ensure the authenticity and integrity of the traffic. However, the implementation of this method imposes a high cost on publishers and ad networks.

\subsection{Countermeasures to Malvertising}
\label{sec:cntmea_Mal}


To avoid malvertising, the authors of~\cite{vratonjic2011online} suggest checking the advertisements regularly and validating their appropriateness. It is the responsibility of the publishers and ad networks to verify the advertising content (whether active or malicious) by performing regular checks. They should avoid publishing advertisements to end-users if publishers and ad networks become aware of any unexpected or unwanted behavior in the code, such as automated redirections. For example, in June 2009, Google launched an investigative research engine to help ad networks by regularly checking the source code of websites. This search engine is publicly available at www.anti-malvertising.com, and enables ad networks to detect potential malvertising providers. Surfers also need to update/install anti-malware programs on their systems to protect against such risks.

The game-theoretic approach~\cite{huang2018bayesian} has been applied to formulate the problem of malvertising, and a mitigation strategy was developed based on data analytics that introduced two Bayesian games between the first player, ad network (defender), and second player, the malvertiser (attacker). The model is not applicable in real-world experiments and has advantages only for advertisers. In~\cite{poornachandran2017demalvertising}, the authors presented a ML-based method for detecting malvertising at publishers' end. They used 15,000 ads and extracted nine features to train Support Vector Machine (SVM) classification. Their results show that 53\% of suspicious ads contain suspicious iFrames. Because the learning-based detection method relies on known attacks patterns, they may not detect unknown attacks.

The authors in~\cite{suresh2019analysis} used the UI-based methodology to detect malware on the mobile operating system. They claimed that it is essential to find the source of the attack in order to detect it. They pointed out that even the main applications may not be malicious, but the web destinations that the user visits can play an important role in spreading the attacks. Thus, they have found three features to have a successful dynamic analytics tool, including triggering the app-web interfaces, detecting malicious content, and identifying the source of an attack. They performed a simulation for two months in two countries to find malware launched through advertisements and web links in applications. Finding the source of the attack helped them detect the responsible entity in the system.

MadDroid~\cite{liu2020maddroid} is a dynamic data analytics framework that has focused on identifying malvertising in mobile environments. The method is based on the characterization of devious ad content that leads to lunch malvertising. The ad content downloaded at runtime from trusted ad networks could serve as a channel for attackers to distribute undesirable contents or even malware. Besides the ad content itself, some unwanted payload may be triggered when the user interacts with the ad content. MadDroid records any network traffic and collects content exchanged between mobile advertisers, ad networks, and user devices. The HTTP hook approach is used to build a mapping between ad libraries and ad hosts. This mapping helps the tool to identify ad traffic from all recorded traffic accurately. Applying MadDroid to 40,000 Android apps, they found that roughly 6\% of apps deliver devious ad contents. Despite the efficiency of the method in finding malware from an ad content perspective, it lacks trustability since the tool and dataset are not publicly available.

%



\section{Lessons Learned and Future Research Directions}
\label{sec:future}

Following the comprehensive survey presented above of the security aspects of online advertising and related techniques, we now summarize the lessons that can be learned and describe possible future challenges and research directions. Some of these have already discussed in previous sections; however, several further challenges and open research issues are dealt with in brief in this section.



\subsection{Lessons Learned}
\label{sec:Lessons}

In many countries, fraudsters have an economic incentive to engage in fraudulent activities and exploit online advertising systems, given the high-risk advertising revenue and the lack of advertising fraud laws. Given that most online services and applications are funded through online advertising revenue interfering with the online advertising business model can have serious consequences. Therefore, there are incentives for stakeholders (e.g., ad networks) to protect advertising revenue and online advertising security. Thus, this work focuses on security issues, which are a crucial element of online advertising.

Despite the rapid growth of online advertising in modern communication networks, there is no comprehensive taxonomy or research documentation to summarize the types of fraud in advertising systems. This paper provided a comprehensive review of fraud activities in online ad systems using a tiered taxonomy to summarize ad fraud at different levels and perspectives. Our paper provides direct answers to key questions such as the main types of fraud in advertising systems, fundamental approaches and features of different types of fraud, and the practical techniques used to detect ad frauds. We provide a detailed description of existing online advertising systems (see Section~\ref{sec:background}) and systems vulnerabilities along with a taxonomy of current attack methods (see Section~\ref{sec:attack_adv}). We have explained who does what and how fraudsters can exploit these vulnerabilities to carry out ad fraud attacks, which we generally divide into three main categories: placement fraud, traffic fraud, and action fraud. For each type of attack, we provide techniques that fraudsters use to profit from ad systems. We discussed the challenges of detecting and reducing advertising fraud and several established countermeasures (see Section~\ref{sec:count_mea}). Table~\ref{table:tbl_conuterm} shows how the various security threats covered in Section~\ref{sec:attack_adv} can be addressed with the available solutions along with the advantages and disadvantages of each.

In particular, we offered various defences to combat click fraud, including the classical approach, the cryptographic approach, and the data analysis approach. We examined the issue of classical criteria, which is related to the lack of efficiency, scalability and maturity. Then, to solve the problems of the classical solutions, we studied different cryptographic methods. We explained that the main idea of these solutions is to change the industry standard to give fraudulent publishers less opportunity to commit fraud. However, the proposed methods suffer from a lack of user privacy due to the sharing of information between third parties and lack of scalability because multiple hash functions are required. Hence, various data analysis techniques have been developed to reduce the problems caused by cryptographic methods. The primary purpose of these technologies is to find specific patterns that identify fake traffic. We have highlighted that the best way to deal with the conflict between the user's privacy and security is to perform statistical analysis on the collected data (such as cookie IDs and IPs) with the help of temporary user identification. The analysis of IPs and cookie IDs is more privacy-friendly than cryptographic methods. Commissioners can track users based on their cookie IDs and IPs. In the current Internet architecture, the use of cookies and IPs to detect fraud can be a less intrusive technique than methods requiring user login. Despite all the defence mechanisms, much research is still needed to design stronger countermeasures and protect the state of e-commerce. It is mainly because the online advertising system has an open platform and needs real-time trading.

The main conclusion stemming from the results reported in this paper is that fraud detection in online transactions is a dynamic field of research, as fraudsters continually invent new techniques for performing fraudulent transactions which seem to be genuine but which cause losses to businesses. Hence, it can be concluded that the real-time detection of ad fraud is the most critical step in making the whole process more efficient. It should also be mentioned that from an economic perspective, search engines need to find as many fraudulent clicks as possible, in order to maximize the ROI for the advertisers. Many efforts are currently being made to mitigate online advertising fraud, since the online advertising market is in a state of expansion and is working to support the needs of advertisers and online commerce providers. Successful fraud management will provide a competitive advantage to ad networks, and will enable them to provide the highest ROI possible to advertisers.

Due to the various limitations on previous investigations and properties of the current online advertising system, we introduce some possible future research directions towards building a reliable, secure, and efficient online advertising ecosystem in Section~\ref{sec:futureDir}. In Section~\ref{sec:securityRes}, we describe some possible solutions to mitigate each open issue.

	\begin{figure}
		\centering
		\includegraphics[width=0.4\textwidth]{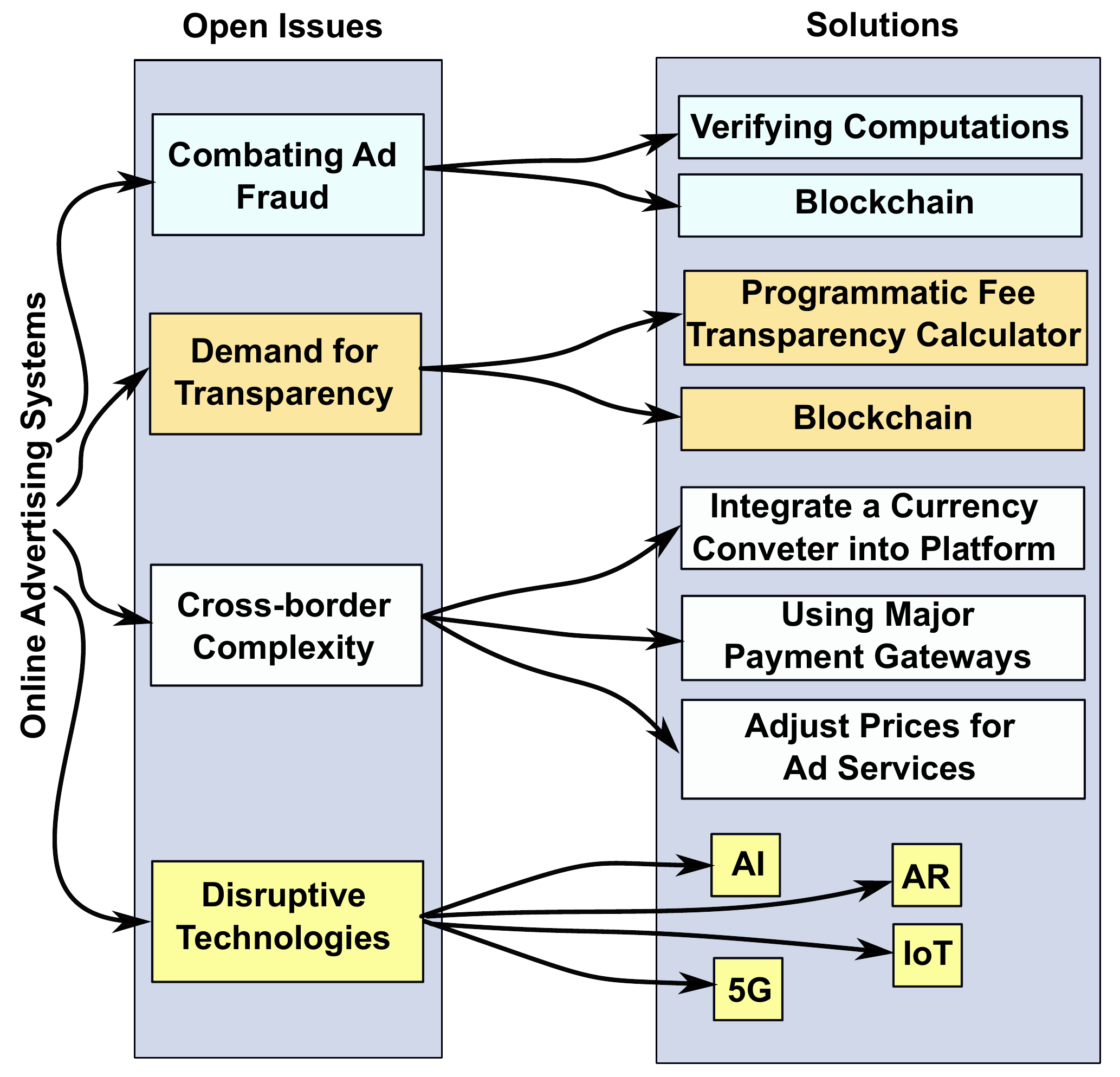}
		\caption{\small Proposed research roadmap for measuring and optimizing the security of online advertising networks. 
			}\label{fig:future}
	\end{figure}

\subsection{Future Direction}
\label{sec:futureDir}

Fig.~\ref{fig:future} shows an overview of four open issues and the corresponding possible solutions. The security, reliability, and efficiency of online advertising systems rely on four major aspects of research, as described below.

\subsubsection{Combating ad fraud}

Although 2021 is expected to be a year of growth, this can be subverted by ad fraud. A report released by Juniper Research states that in 2018, about \$42 billion was lost to ad fraud in the online advertising business~\cite{Juniper}. It is expected that this amount will grow to \$100 billion by 2023~\cite{Juniper}. The damages do not simply involve financial loss, and can affect user privacy and hide the best performing marketing channels. To deal with these damages, growth marketers must consider fraud prevention as a priority. The report claims that attackers tend to apply methods such as domain spoofing to increase the number of clicks by misrepresenting a low-quality site to resemble a high-quality website, rather than using techniques such as app install farms. As a result, it is essential to detect which ad clicks are fake and which are genuine, not an easy task in real-time bidding.

\subsubsection{Demand for transparency}

The report in~\cite{Nielsen} points out that the majority of the cost allocated to online advertising currently goes directly to waste, due to fraud or off-target audiences. However, there are ways to adapt, and transparency can play a significant role in this. For the entities that are involved in the ad industry, it is vital to know where their banners are served and where their budgets are spent, since if control over the budget allocated to the ad campaign is lost, advertisers will not know what has been spent where. Advertisers and publishers are doing business, and their activities therefore aim to make money, but the fragmentation of this economy means that media customers spend more high-priced than it's worth.

\subsubsection{Cross-border complexity}

This aspect aims to attract and protect global users who require multi-currency pricing options. For example, customers from all parts of the world trust ad providers to give them ad services. However, the payment methods by ad providers are not acceptable. As a result, to gain customer loyalty, ad service providers need to allow them to change money on their side at suitable exchange rates~\cite{Mcurren}. In this way, they can build a sustainable and secure platform to execute different multi-currency scenarios.

\subsubsection{Disruptive technologies}

The online advertising industry has been significantly penetrated by technological innovations like the Internet of Things (IoT)~\cite{aksu2018advertising}, artificial intelligence (AI)~\cite{choi2020identifying}, augmented reality (AR)~\cite{tsai2020inspection}, and 5\textsuperscript{th} generation mobile Internet (or 5G)~\cite{zhao2020imagination}. In 2018, for example, Google launched a beta experiment involving automatic ad placement on the basis of AI, and publishers' incomes increased by 10\%~\cite{AutoAds}. To gain a competitive advantage in the market to survive, an enterprise needs to adapt to these changes faster than others, and the future of companies who are not ready for the newest technologies is in question.

\subsection{Suggestion of Security Responses}
\label{sec:securityRes}

In this section, we propose some responses to the challenges introduced in Section~\ref{sec:futureDir}. 

\subsubsection{Responses for ad fraud}
\label{sec:securityResfraud}

Ad fraud has become a significant concern for everyone involved in the ad industry, and can lead to reductions in trustworthiness and campaign effectiveness, and the siphoning of budgets. 

The industry's primary solution for combating all types of fraud is the use of ML to analyze the history of attacks and how they appeared, to help companies predict what will happen next~\cite{essF}. One of the best and most efficient solutions to prevent ad fraud is to apply sophisticated click validation mechanisms. This increases the workload for fraudsters aiming to steal advertisers' and brands' budgets, and makes it uneconomical for them. In 2019, Adjust~\cite{Adjust} proposed a standard based on click validation in which ad channels send impressions with a unique identifier before the click claim is sent.

As mentioned previously, whenever users click on a hyperlink in a publisher's website, the advertiser must pay a fee. The question therefore arises as to how an advertiser can verify that the bill received from the publisher is correct. This poses a challenge and remains an open issue. In this case, our suggestion is to apply \emph{Verifying Computations} without requiring the user to re-execute~\cite{walfish2015verifying} them. The fundamental theorem behind this is a probabilistic proof system, which is composed of two elements, a prover and a verifier. The prover aims to prove a mathematical assertion (so-called proof) for the verifier, while the verifier checks the proof.

However, in practice, this computational technique is not economically sound. We, therefore, propose the use of a blockchain-based scheme for validation and verification. Using blockchain-based solutions to tackle ad fraud also studied by Kshetri et al.~\cite{kshetri2019online}. The concept underlying the blockchain is Distributed Ledger Technology (DLT), which helps various untrusting and distributed agents to transmit data in a trusted, secure, and valid way by providing distributed validation, transparency, and cryptographic immutability. DLT is a technology that starts with the cryptocurrency of bitcoin, followed by smart contracts. The properties of DLTs, such as consensus protocols, provenance and immutability, are used in the model for secure transactions~\cite{asante2021distributed}. The consensus protocol is that a network must approve all transactions of all players. Validation algorithms vary from system to system, but the bottom line is that the system only accepts transactions that are made under valid rules. Provenance means knowing the source of the data. Immutability can be defined as the ability to keep data recorded in the DLT unchanged. The problem of trust between entities that do not trust each other can be solved with these three DLT features. Moreover, this is very well in line with the lack of trust in the online advertising ecosystem. The pervasiveness of ad fraud is due to the lack of an intermediary that can track online advertising to increase trust and reduce some concerns. Blockchain transparency helps detect fraudulent traffic and improves ad delivery. With the ability to import transparency to the system and track assets online, blockchain can help the advertising system to reduce ad fraud, if not stop it completely. It is possible to know who did what and when. Recently, a wide range of applications (such as healthcare~\cite{shae2018transform} and genomics~\cite{ozercan2018realizing}) have begun to use the blockchain to guarantee trustworthiness in interactions among untrusting agents. Thus, the blockchain is an appropriate mechanism to ensure trust in cases that require long-running computations. We believe that an important future research direction in using validation clicks to fight ad fraud could be to investigate how blockchain-based validation can be extended and used to ensure effective, trusted verifiable computations.

\subsubsection{Responses for transparency demand}

High levels of transparency play a significant role in building trust between entities in the online advertising system and customers~\cite{trans, portes2020should}. This also affects the relationship between the publisher and advertiser. One way to help bring transparency over cost is to create a real-time analytic method to follow all activities. In the following, we highlight some other technologies and tools that can improve and guarantee transparency.

\begin{itemize}
	
\item In 2016, IAB released a \emph{Programmatic Fee Transparency Calculator} to add transparency to the collaboration between publisher and advertiser~\cite{IAB}. This tool was designed to help actors in the online advertising market to define and apply cost models differently. In this way, they have the flexibility to enter their planning rates and budgets into the calculator, and then select the available advertising technologies for the campaign. It is essential to mention that the calculation cost model is based on the ``\% of media.'' 

\item The blockchain can provide security and transparency for the transfer of data from advertiser to publisher~\cite{blocktran}. It is also possible to do real-time transactions by exploiting blockchain technology, especially when the price is obvious to all participating members of the supply chain. Blockchain transparency can improve the ad delivery process. Advertisers can combine ad data with data provided by ad viewers to increase the effectiveness of their ads. Blockchain has the potential to change the way online ads are paid, sold and measured. It allows advertisers to see if their ads are being delivered or reaching the right customers. Advertisers can track who opened the ad and where is the customer location. A few advertising tools are available to cope with the transparency challenge, including Havas~\cite{Havas} and Apomaya. These platforms aim to support transparency by calculating the fees that media buyers must pay.

\end{itemize}

\subsubsection{Responses for cross-border complexity}

Engineers have expertise in developing ad software that facilitates multiple-currency and cross-border operations. They are aware of how to create and maintain smart billing services that can support multi-currency payments. We identify some other techniques for coping with the challenges of cross-border complexity as follows.

\begin{itemize}
	
	\item One technique is to integrate a currency converter calculator into a pre-built framework. This requires finding an Application Programming Interface (API), such as currencylayer, Fixer, or XE Currency Data, to allow regular updating of currency exchange rates and access to the maximum number of worldwide currencies.
	
	\item Another technique is to provide customers with access to different payment gateways, including PayPal, Secure-Pay, Stripe, Authorize.Net, etc. Offering diverse payment options can help to attract and retain loyal customers~\cite{Small}. The new digital currency, called Bitcoin, as a peer-to-peer electronic cash system can also help the payment process. This cryptocurrency mechanism can not only solve the double-spending problem~\cite{treiblmaier2019combining}, but it sets out a new paradigm for performing transactions and exchanging value in an online environment. The interactive and universal features of bitcoin allow marketers to bypass intermediaries, transmit their business content, and reduce costs. For example, retailers typically pay 3\% of payment processing to credit card companies, and many online platforms receive sales commissions. Blockchain technology can help brands limit or remove costs and eliminate worthless activities in the middle layer. Thus, the technology can potentially extend the direct relationship between brands and consumers.
	
	\item Ad services can also be provided with adjustable prices by considering the average transaction cost across a specific country, since a given amount might be adequate for one country but too high for another.
	
\end{itemize}

\subsubsection{Responses for Disruptive technologies}

It is not an easy task to apply cutting-edge technologies when the traditional types work well. For example, it is difficult for an advertiser to change their ad campaigns to the emerging ones. However, in this new era, there is a need to adapt and be aware of the latest technologies, and the domain of online advertising systems is no exception. Emerging technologies such as AI, AR, IoT, and 5G can help ad tech companies in several ways~\cite{camilleri2020use}. For example, the role of AI is three-fold. Firstly, the use of AI-based chatbot applications will motivate users to buy products, since a chatbot allows them to ask questions, give commands and receive services in a conversational style~\cite{adam2020ai}. An AI chatbot can read data, analyze complex information and make decisions based on this information. Depending on the customer's question, the system should refer them to a specific social group to demonstrate the items that can be purchased. Secondly, AI provides a method of targeted advertising. Assisted by the application of ML algorithms to big data, AI can automatically sort marketing messages and deliver them to the target users, making ad targeting more accurate and cost-effective~\cite{zhang2020deep}. Thirdly, running AI-based algorithms allows ad mediation to be optimized to maximize profits for publishers by finding the best-matched slots for their advertisements. AI-based advertising helps companies in four ways: $(i)$ by displaying personalized advertisements to the relevant customers and minimizing human effort; $(ii)$ by interacting with audiences in a natural way; $(iii)$ by reducing errors using a data-oriented approach for network selection; and $(iv)$ by saving time through automating the process of ad publishing.

Although AR-based marketing is in its infancy, it has become interesting to marketers. Since everyone has a smartphone, advertisement based on AR is now much easier than before. For example, stores can install AR-driven ad applications to send customers popup advertisements to tell them about products, and consequently attract customers to purchase items. AR-based advertising helps companies in three ways: $(i)$ by providing targeted and innovative contextual advertising; $(ii)$ by improving customer experience and making it unique and immersive; and $(iii)$ by boosting customer loyalty through interactive advertising.

A report from the IAB found that around 65\% of people in the US own at least one IoT device, and are interested in receiving advertisements on IoT screens~\cite{IoT}. IoT technologies can therefore provide new levels of ad targeting~\cite{aksu2018advertising}. IoT data can be used to dig even deeper into customers' habits, interests, preferences, and other factors, and allows advertisers to learn more about their customers to create customer personas and targeted ad campaigns. It is also possible to integrate cloud solutions with various gateways to achieve better results in the ad campaign. IoT-based ad software can help advertising companies in three ways: $(i)$ better recognition and prediction of consumers' individual preferences and needs to increase the efficiency and accuracy of target advertisements; $(ii)$ increased user engagement and satisfaction by providing them with valuable information about products; and $(iii)$ improved ad campaign effectiveness.



The arrival of 5G promises to open up substantial new opportunities to advertisers and customers. The possibility of achieving Internet speeds 20 times faster than 4G is an enticing one for both of them. The faster network speeds and high-resolution screens through 5G will allow video resolution to be increased and page loading times to be reduced, creating more interaction between customers and advertisers. These features lead to more customer engagement for e-commerce activities, more time spent on e-commerce websites and more online shopping~\cite{kshetri20185g}. Needless to say, to fully exploit the potential of 5G in the advertising industry, all the entities in the industry should prepare themselves before launching 5G. In the following, we identify some of the issues that should be considered.

\begin{itemize}[leftmargin=*]
	
	\item \textbf{Faster load speeds.} Despite the advent of new technologies, like AI, AR, and 3D modeling, which have revolutionized the ad market, advertisers may not be attracted to online advertising due to issues relating to speed. With the high speeds of 5G, a new era will open up for advertisers to exploit customer profiling, ad creative, targeting, and many more aspects. 5G will increase the speed of a device from 45 Mbps up to a maximum of one gigabit, meaning that response times will be a few milliseconds, thus leading to a decrease in latency~\cite{sung20195g}. This can provide a better space for the use of streaming video (or even deeper augmented and virtual reality) to create advertisements~\cite{D5G}. StateFarm reported a 500\% increase in mobile ad click-through rates due to its virtual reality ads, indicating the potential of virtual reality ads to grab the audience's attention \cite{lee2021adcube}. It also opens the way for creating video advertisements, giving customers the chance to stop scrolling the web page to watch high-resolution advertisements.
	

	\item \textbf{Reach across channels.} The efficiency of the online advertising ecosystem depends on the communication technology infrastructure~\cite{helberger2020macro}. Speed and access to Internet connections affect the advertising industry's ability to reach consumers. With faster speeds everywhere, more devices will be connected and participate in the online advertising world. Hence, with the introduction of 5G networks, consumers will eventually experience much faster wireless connectivity. 5G can not only help with ad creation but will also help advertisers reach audiences more effectively~\cite{5G}. The targeting of audiences with low-speed Internet was not straightforward, but 5G will pave the way for creating a range of channels that will enable advertisers to connect directly with consumers.
		
	
	\item \textbf{Unlocking identity.} It is worth pointing out that advertisers with more digital touchpoints are more likely to be selected by customers, and should consider this a chance to learn more information about their audiences. Not surprisingly, the most significant impact of 5G will be on the quality and quantity of data in the system, which will allow the advertiser to target and capture the correct audience more effectively~\cite{kshetri20185g}. To achieve maximum benefit from the new data, it is vital to make sure all the basis includes the right technology partners, the right infrastructure and the right kind of privacy measures for the use of the data. Moreover, in the run-up to the introduction of 5G, marketers should warn stakeholders to consider the principles of privacy by design when using this valuable data.
	
\end{itemize}


The disruptive technologies that have been developed over the last decade promise new solutions and new ways to connect with customers and markets; however, as with any opportunity for growth, these developments are not without risks, and businesses should start considering these now. For example, predictions made by Ericsson show that 29 billion devices will be connected to the IoT by 2022~\cite{risk}. This growth in the number of IoT connections will lead to increased security threats. Hence, markets and businesses need to ensure that their IoT-connected devices are safe, that default passwords are not used and that all security updates are installed.

\section{Conclusion}
\label{sec:coclu}

Online advertising is vital in sustaining the economy of the Internet, since each party in the system can gain profit. However, in many countries, there is a lack of legal protection against ad fraud, and given the amount of ad revenue at stake, online advertising has become a target for criminals to gain financial incentives through fraudulent activities.

In this article, we have investigated and discussed the security aspects of the online advertising market. We first gave a brief introduction to the online advertising system, followed by the fundamental concepts that have emerged in relation to the online advertising system. Next, we presented a state-of-the-art study of the various forms of security attacks on the system that arise from the weaknesses of the ecosystem. We then proposed a comprehensive taxonomy of ad fraud to describe these threats in global terms and facilitate cooperation among research hyper refers to deal with ad fraud attacks. We classified the existing solutions that have been proposed in the literature to cope with these attacks, along with the limitations and effectiveness of these solutions. Finally, we presented our view of current research challenges and future directions to improve existing security solutions in the online advertising system.

\bibliographystyle{IEEEtran}
\bibliography{Bibliografia}
\vspace{-35px}
\begin{IEEEbiography}[{\includegraphics[width=1in,height=1.25in,clip,keepaspectratio]{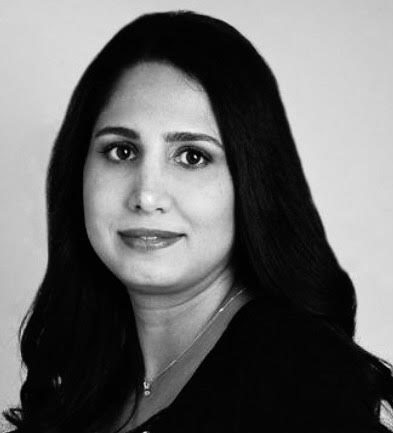}}]{Zahra Pooranian}\textbf{(M'17-SM'21)} is a postdoctoral fellow in Network Security at the 5G \& 6G Innovation Centre (5GIC \& 6GIC), Institute for Communication Systems (ICS), University of Surrey, Guildford, UK. Before joining the University of Surrey, she was postdoctoral at The Alan Turing Institute, London, UK. She was working on Enhancing Security and Privacy of National Identity Systems project. From 2017 to 2019, she was postdoctoral in Network Security at the University of Padua, Padua, Italy. She received her Ph.D. degree in Computer Science Sapienza University of Rome, Italy, in February 2017. She is a (co)author of several peer-reviewed publications (h-index=20, citations=1136+) in well-known conferences and journals. She is an Editor of KSSI transaction on Internet and information systems and Future Internet. Her current research focuses on Machine Learning, Social Network Security, Cloud Security and Cloud/Fog Computing. She was a programmer in several companies in Iran from 2009-2014, respectively. She is a Senior Member of IEEE. For additional information: \url{https://zahrapooranian.github.io/Zahra/} 
\end{IEEEbiography}

\vspace{-35px}
\begin{IEEEbiography}[{\includegraphics[width=1in,height=1.25in,clip,keepaspectratio]{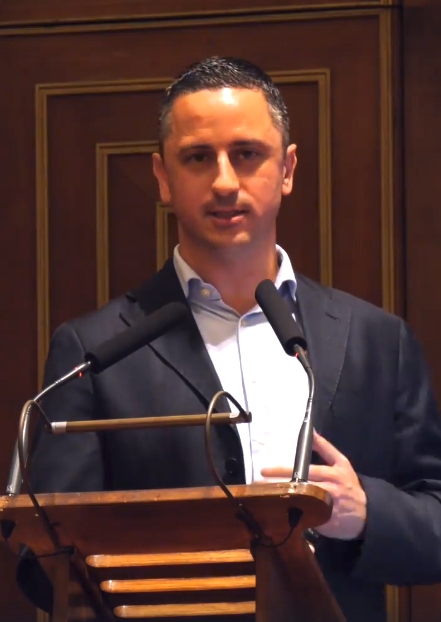}}]{Mauro Conti} is Full Professor at the University of Padua, Italy. He is also affiliated with TU Delft and University of Washington, Seattle. He obtained his Ph.D. from Sapienza University of Rome, Italy, in 2009. After his Ph.D., he was a Post-Doc Researcher at Vrije Universiteit Amsterdam, The Netherlands. In 2011 he joined as Assistant Professor the University of Padua, where he became Associate Professor in 2015, and Full Professor in 2018. He has been Visiting Researcher at GMU, UCLA, UCI, TU Darmstadt, UF, and FIU. He has been awarded with a Marie Curie Fellowship (2012) by the European Commission, and with a Fellowship by the German DAAD (2013). His research is also funded by companies, including Cisco, Intel, and Huawei. His main research interest is in the area of Security and Privacy. In this area, he published more than 400 papers in topmost international peer-reviewed journals and conferences. He is Area Editor-in-Chief for IEEE Communications Surveys \& Tutorials, and has been Associate Editor for several journals, including IEEE Communications Surveys \& Tutorials, IEEE Transactions on Dependable and Secure Computing, IEEE Transactions on Information Forensics and Security, and IEEE Transactions on Network and Service Management. He was Program Chair for TRUST 2015, ICISS 2016, WiSec 2017, ACNS 2020, CANS 2021, and General Chair for SecureComm 2012, SACMAT 2013, NSS 2021 and ACNS 2022. He is Senior Member of the IEEE and ACM. He is a member of the Blockchain Expert Panel of the Italian Government. He is Fellow of the Young Academy of Europe. From 2020, he is Head of Studies of the Master Degree in Cybersecurity at University of Padua. For additional information: \url{ http://www.math.unipd.it/~conti/}
\end{IEEEbiography}\vspace{-35px}

\begin{IEEEbiography}[{\includegraphics[width=1in,height=1.25in,clip,keepaspectratio]{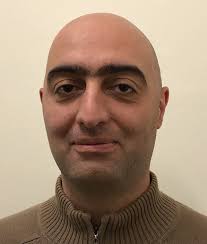}}]{Hamed Haddadi} is a Reader in Human-Centred Systems and the Director of Postgraduate Studies at the Dyson School of Design Engineering at The Faculty of Engineering, Imperial College London. He serves as a Security Science Fellow of the Institute for Security Science and Technology, and is an Academic Fellow of the Data Science Institute. In his industrial role, he is a Visiting Professor at Brave Software where he works on developing privacy-preserving analytics protocols. He enjoys designing and building systems that enable better use of our digital footprint, while respecting users' privacy. For additional information: \url{https://www.imperial.ac.uk/people/h.haddadi}
\end{IEEEbiography}\vspace{-35px}

\begin{IEEEbiography}[{\includegraphics[width=1in,height=1.25in,clip,keepaspectratio]{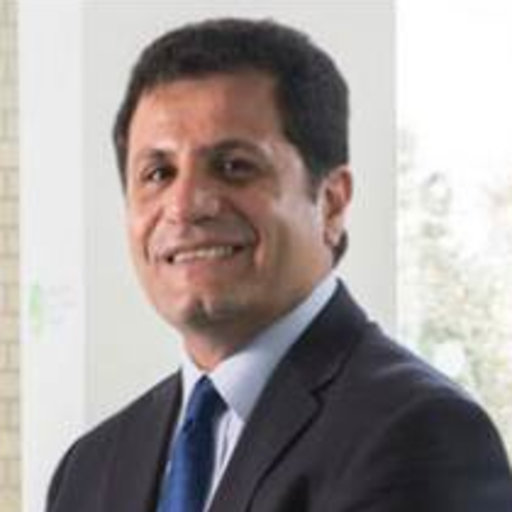}}]{Rahim Tafazolli} \textbf{(SM'09)} is a professor and the Director of the Institute for Communication Systems (ICS) and 5G Innovation Centre (5GIC), the University of Surrey in the UK. He has over 30 years of experience in digital communications research and teaching. He has published more than 500 research papers in refereed journals, international conferences and as invited speaker. He is the editor of two books on ``Technologies for Wireless Future'' published by Wiley Vol.1 in 2004 and Vol.2 2006. He is co-inventor on more than 30 granted patents, all in the field of digital communications. He was appointed as Fellow of WWRF (Wireless World Research Forum) in April 2011, in recognition of his personal contribution to the wireless world. As well as heading one of Europa's leading research groups. He is regularly invited by governments to advise on network and 5G technologies and was advisor to the Mayor of London with regard to the London Infrastructure Investment 2050 Plan during May and June 2014. For more information: \url{https://www.surrey.ac.uk/people/rahim-tafazolli}
\end{IEEEbiography}

\end{document}